\documentclass[12pt]{iopart}

\newcommand{\commento}[1]{}
\newcommand{\eqref}[1]{(\ref{#1})}
\usepackage{iopams}  
\usepackage{graphicx}
\usepackage{amsbsy}
\usepackage{amsfonts}
\usepackage{paralist}
\usepackage{color}

\newcommand{\ebX}{\epsilon {\bf X}}
\newcommand{\bX}{\textbf{X}}

\begin{document}

\definecolor{gray}{RGB}{159,157,157}

\title[Two temperatures]{Non-equilibrium and information: the role of cross-correlations}

\author{A. Crisanti$^1$, A. Puglisi$^1$ and D. Villamaina$^2$}
\address{$^{1}$ CNR-ISC and Dipartimento di Fisica, Universit\`a Sapienza - p.le A. Moro 2, 00185, Roma, Italy}
\address{$^{2} $ Univ. Paris-Sud \& CNRS, LPTMS, UMR 8626, Orsay 91405 France}



\begin{abstract}
  We discuss the relevance of information contained in
  cross-correlations among different degrees of freedom, which is
  crucial in non-equilibrium systems. In particular we consider a
  stochastic system where two degrees of freedom $X_1$ and $X_2$ - in
  contact with two different thermostats - are coupled together. The
  production of entropy and the violation of equilibrium
  fluctuation-dissipation theorem (FDT) are both related to the
  cross-correlation between $X_1$ and $X_2$. Information about such
  cross-correlation may be lost when single-variable reduced models,
  for $X_1$, are considered. Two different procedures are typically
  applied: (a) one totally ignores the coupling with $X_2$; (b) one
  models the effect of $X_2$ as an average memory effect, obtaining a
  generalized Langevin equation.  In case (a) discrepancies between
  the system and the model appear both in entropy production and
  linear response; the latter can be exploited to define effective
  temperatures, but those are meaningful only when time-scales are
  well separated. In case (b) linear response of the model well
  reproduces that of the system; however the loss of information is
  reflected in a loss of entropy production. When only linear forces
  are present, such a reduction is dramatic and makes the average
  entropy production vanish, posing problems in interpreting FDT
  violations.
\end{abstract}

\maketitle

\section{Introduction}

Energy and information are well known to be related: the conceptual
Maxwell's demon experiment is a popular representation of such an
empirical fact. Extracting energy from a cold to a hot reservoir
requires a device able to discern fast molecules from slow ones,
i.e. requires the processing of information. Proofs in simplified
models and overwhelming experimental evidence leads to the conclusion
that this information costs, in energy, more than what is gained in
the extraction~\cite{M71,S29,L61}. Such an issue is recently
receiving a renewed interest, both theoretical~\cite{SU10,AS12} and
experimental~\cite{C12}, in the contest of small systems and
non-equilibrium thermodynamics.

What is important, in this debate, is a proper evaluation of the
information needed to perform the physical process under
examination. This in turn amounts to have a good model of the system:
in particular it is crucial that the model correctly reproduces the
information fluxes involved in its dynamics. Such an issue appears to be
delicate, since modelling implies some level of coarse-graining
and, as a consequence, a {\em loss} of information~\cite{KPB07,RJ07,PPRV10}.

Remaining in the framework, mentioned above, of energy flowing between
two different - and disconnected - reservoirs, an interesting example
is provided by the information associated to the energy flux, which is
measured in the following way. We consider two probes,
e.g. two colloids or big molecules, which are coupled by a linear
spring. Each probe is in contact with one of the two reservoirs, so
that the spring between the probes also connects the thermostats. The
interaction between a probe and its own thermostat - described in
details in the text below - is characterized by a typical time which
is in principle different for each probe~\cite{CK00,VBPV09,ZBCK05}. At equilibrium (identical
thermostats) such typical times are not relevant, but they become
important in the more general non-equilibrium case.

In such a system it is straightforward to compute the so-called entropy
production rate, which is a measure of how fast information is created
in the ensemble of probes' pairs or, equivalently, of how fast this
ensemble would relax to equilibrium if allowed to. The relaxation to
equilibrium is forbidden by some (undetailed) external
constraint which prevents the two thermostats from equilibrating and
which continuously dissipates the information so far created, allowing
the system to achieve a non-equilibrium steady state~\cite{SVGP10} (see also~\cite{V06,FI11}). Of course, if
the reservoirs have the same temperature, this entropy production rate
vanishes and the steady state satisfies detailed balance~\cite{PV09}.

In this paper we discuss the effect of modelling the system by removing
from the description one of the two probes. Two cases
are interesting: (a) one simply ignores the existence of the coupling
with the second degree of freedom; (b) one keeps some information
about such a coupling, but - for the purpose of making things simpler
- replaces it with proper memory terms and an effective colored noise,
resulting in a generalized Langevin equation~\cite{KTH91} with the
second kind fluctuation dissipation relation which is not
satisfied. In case (a), one expects equilibrium, therefore even simple
observations (for instance linear response) do not agree with
expectations, and the departure of such agreement may be interpreted
to define non-equilibrium effective temperatures~\cite{CKP97,CR03}. However such a
procedure is really meaningful only in the presence of strong
separation of time-scales, otherwise unphysical effective temperatures
appear~\cite{VBPV09}. In case (b) the statistics of the dynamics of the
remaining probe is properly reproduced, including linear
response. However the measure of the rate of information creation
(entropy production) is underestimated. This discrepancy becomes dramatic in
the case of linear couplings: in that case entropy production
completely vanishes, resulting in the idea that the system is at
equilibrium.

The plan of the paper is the following. In Section~\ref{sec:models} we
present the system with two variables, justifying it from a physical
point of view, and we offer a review of its statistical dynamics
properties. The system is taken fully linear from the beginning, but -
as also detailed in the Appendixes - most of the results are more
general.  The importance of cross-correlation between the two
variables, and the effect of removing (in different ways) the second
one from the description of the system, is discussed in
Section~\ref{sec:entropy_information}. Finally in
Section~\ref{sec:conclusion} we put our result in a more general
perspective, discussing the role of {\em channels} for the transport
of energy and information and how they depend on the chosen level of
description. 

Appendixes contain not only lengthy calculations accompanying the main
results of the paper, but also deeper insights into the problem:
~\ref{appA} discusses also a partially non-linear case, as well as
formulations in (time) Fourier space; \ref{appinertia} discusses the case of
the same system with inertia, such that one of the degree of freedom
has different parity under time-reversal; \ref{appB} explains the
subtle conditions necessary to reduce the system with two variables to
the model with one variable and memory; finally in \ref{example} we offer
an explicit example where the entropy production in the full description
(two Markovian variables) has an additional contribution, with respect
to the reduced description (one variable with memory), which carries
crucial information about the difference of temperature among the two
thermostats.

\section{A system with two temperatures} \label{sec:models}

Most of the ideas in this paper are illustrated by using a simple
simple stochastic non-equilibrium system with two coupled degrees of
freedom. The purpose of this section is to describe it and recall the
main known properties of its dynamics.
Our system is described by two coupled Langevin equations:
\begin{eqnarray} \label{eqmotion}
\dot{X_1}&=& -\alpha X_1+\lambda X_2 + \sqrt{2D_1}\phi_{1} \nonumber \\
\dot{X_2}&=& -\gamma X_2 +\mu X_1 + \sqrt{2D_2}\phi_{2} 
\end{eqnarray}
where $\phi_{1}$ and $\phi_{2}$ are uncorrelated white noises, with
zero mean and unitary variance.

\begin{figure}[h!]
\includegraphics[clip=true,width=0.5\columnwidth]{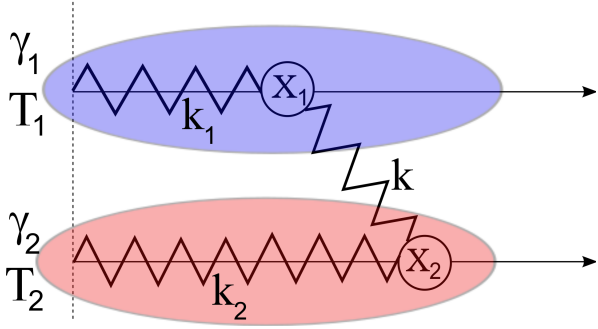}
\caption{graphical representation of the system described by (\ref{PhysicalInterpretation}). \label{fig1}} 
\end{figure}

The above stochastic equations can be thought as modelling the system
portrayed in Fig.~\ref{fig1}. The system includes two particles (for
simplicity in one dimension), with positions $x_1$ and $x_2$ and
momenta $p_1$ and $p_2$ whose Hamiltonian is given by
\begin{equation}
\mathcal{H}_{tot}=\frac{p^{2}_{1}}{2m_{1}}+\frac{p^{2}_{2}}{2m_{2}}+\frac{1}{2}k_1
x_1^{2}+\frac{1}{2}k_2 x_2^{2}+\frac{1}{2}k (x_1-x_2)^{2}\label{Hamiltonian}.
\end{equation}

Each particle $i$ is moving in a dilute fluid which exerts a viscous
drag with coefficient $\gamma_i$, and which is coupled to a thermostat
with temperature $T_i$; a natural way of modelling for the dynamics of this
system is the following:
\begin{eqnarray} \label{complete}
\dot{p_{1}} &=& -\frac{\partial \mathcal{H}}{\partial x_1}-\gamma_{1}\dot{x_1}+\sqrt{2\gamma_{1}T_{1}}\phi_{1} \nonumber \\
\dot{p_{2}} &=& -\frac{\partial \mathcal{H}}{\partial x_2}-\gamma_{2}\dot{x_2}+\sqrt{2\gamma_{2}T_{2}}\phi_{2}.
\end{eqnarray}

Now, by taking the overdamped limit we get:
\begin{eqnarray}
\gamma_{1}\dot{{x_1}} &=& -(k_1+k)x_1+k x_2+\sqrt{2\gamma_{1}T_{1}}\phi_{1} \nonumber\\
\gamma_{2}\dot{{x_2}}&=& kx_1-(k+k_2)x_2 +\sqrt{2\gamma_{2}T_{2}}\phi_{2} \label{PhysicalInterpretation}
\end{eqnarray}
which corresponds to model~(\ref{eqmotion}) by identifying $X_1 \to x_1$, $X_2 \to x_2$ and
\begin{equation} \label{phys_ident}
\alpha \to \frac{k_1+k}{\gamma_1} \;\;\;\; \lambda \to
\frac{k}{\gamma_1} \;\;\;\; \gamma \to \frac{k+k_2}{\gamma_2} \;\;\;\;
\mu \to \frac{k}{\gamma_2} \;\;\;\; D_1 \to \frac{T_1}{\gamma_1}
\;\;\;\; D_2 \to \frac{T_2}{\gamma_2}.
\end{equation}

\subsection{Steady state properties}
System~\eqref{eqmotion}, in a more compact form, reads
\begin{equation}
\frac{d\bf X}{dt}=-A{\bf X}+ \mbox{\boldmath$\phi$}, \label{2main}
\end{equation}
where ${\bf X}\equiv(X_1,X_2)$ e $\mbox{\boldmath$\phi$}\equiv(\phi_1,\phi_2)$ are
2-dimensional vectors and $A$ is a real $2\times 2$ matrix, in general
not symmetric, $\mbox{\boldmath$\phi$}(t)$ is a Gaussian
process, with covariance matrix:
\begin{equation}
\left<\phi_{i}(t')\phi_{j}(t)\right>=2D_{ij}\delta(t-t'),
\end{equation}
and 
\begin{equation}
A=\left(
\begin{array}{cc}
\alpha & -\lambda  \\
 -\mu  & \gamma 
\end{array}
\right)\phantom{mmmmm}
D=\left(
\begin{array}{cc}
  D_{1} & 0  \\
0  & D_{2} 
\end{array}
\right)
\end{equation}
In order to reach a steady state, the real parts of $A$'s eigenvalues
must be positive. This condition is verified if $\alpha+\gamma >0$
and $\alpha\gamma-\lambda\mu>0$. Extension to a generic dimension $d$
and non-diagonal matrices $D$ (which however must remain symmetric) is
straightforward.

The steady state is characterized by a bivariate Gaussian
distribution~\cite{R89}:
\begin{equation}
\rho({\bf X})=N\exp{\left(-\frac{1}{2}{\bf X}\sigma^{-1}{\bf
    X}\right)} \label{GaussNESS}
\end{equation}
where $N$ is a normalization coefficient and the matrix of covariances
$\sigma$ satisfies 
\begin{equation} \label{ness}
D=\frac{A\sigma+\sigma A^{T}}{2}.
\end{equation}
Solving this equation gives
\begin{equation}
\sigma=\left(
\begin{array}{cc}
 \frac{D_2 \lambda ^2-D_1 \mu  \lambda +D_1 \gamma  (\alpha+\gamma )}{(\alpha+\gamma ) (\alpha \gamma -\lambda  \mu )} & \frac{D_2 \alpha
   \lambda +D_1 \gamma  \mu }{(\alpha+\gamma ) (\alpha \gamma -\lambda  \mu )} \\
 \frac{D_2 \alpha \lambda +D_1 \gamma  \mu }{(\alpha+\gamma ) (\alpha \gamma -\lambda  \mu )} & \frac{D_1 \mu ^2-D_2 \lambda  \mu
   +D_2 \alpha (\alpha+\gamma )}{(\alpha+\gamma ) (\alpha \gamma -\lambda  \mu )}
\end{array}
\right).\label{sigmamatrix}
\end{equation}

Moreover, in this system it is also possible to calculate the path
probabilities. The probability of a trajectory $\{ {\bf X}(s)\}_0^t$
in the phase space can be written in the following form:
\begin{equation}
 P(\{ {\bf X}(s)\}_0^t)=\int \mathcal{D}{\bf \phi} P({\bf \phi})\delta(\dot{{\bf X}}+A{\bf X}-\phi),
 \end{equation}
where the integral involves all the possible realizations of the noise with the corresponding weight. By introducing auxiliary variables, using the integral representation
of the delta function, one obtains~\cite{MSR73}:
\begin{equation}
 P(\{ {\bf X}(s)\}_0^t)\sim\int \mathcal{D}{\bf \hat{X}} e^{S({\bf X},\hat{\bf X})}
\label{MartinSiggiaRose}
 \end{equation}
Where $S({\bf X},\hat{\bf X})=\frac{1}{2} \hat{{ \bf X}} D \hat{{ \bf
    X}}+i\hat{{\bf X}}(\dot{{\bf X}}+A{\bf X})$.\\ In the following,
we will also use the Onsager-Machlup expression for the path
probabilities, which is also obtained by integrating expression
(\ref{MartinSiggiaRose}) over the \emph{hat} variables~\cite{OM53}
\begin{equation}
P(\{ {\bf X}(s)\}_0^t)\sim e^{\left((\dot{{\bf X}}+A{\bf X})D^{-1}(\dot{{\bf X}}+A{\bf X})\right)} \label{OnsagerMachlup}
 \end{equation}
Expression (\ref{OnsagerMachlup}) has the advantage of not needing the
presence of auxiliary fields.

Equilibrium is defined as the regime where path and their
time-reversal have the same probability, i.e.
\begin{equation}
\rho[{\bf X}(0)]P(\{ {\bf X}(s)\}_0^t) = \rho[\mathcal{I}{\bf X}(t)] P(\{\mathcal{I}{\bf X}(s)\}_0^t),
\end{equation}
where $\overline{{\bf X}}$ is the time reversed phase point, and
$\rho$, defined in~(\ref{GaussNESS}), represent the probability of the initial
condition.  It is easy to verify that such a condition leads to
\begin{eqnarray} \label{eqC}
C_{ij}(t)&=\epsilon_i \epsilon_j C_{ji}(t),\\
\dot{C}_{ij}(t)&=\epsilon_i \epsilon_j \dot{C}_{ji}(t),
\end{eqnarray}
where we have defined, for $t\ge 0$, $C_{ij}(t)\equiv \langle
X_{i}(t)X_{j}(0)\rangle$ the time-delayed cross-correlation and
$\epsilon_i$ the parity ($1$ or $-1$) under time-reversal of the $i$-th
variable. 
Considering that one has for the matrix of time-delayed correlations~\cite{R89}
\begin{eqnarray}
C(t)&=e^{At}\sigma \nonumber\\
\dot{C}(t)&=e^{At}A\sigma \label{CorrOnsager},
\end{eqnarray}
by evaluating the above conditions at $t=0$, it is seen that the
equilibrium definition~\eqref{eqC} leads to two important conditions:
\begin{enumerate}

\item $\sigma_{ij}=0$ if $\epsilon_i \neq \epsilon_j$ (because $\sigma$ is symmetric by construction);

\item $(A\sigma)_{ij}=\epsilon_i \epsilon_j (A\sigma)_{ji}$, which are
  the so-called {\em Onsager reciprocal relations}, being $A\sigma$
  the Onsager matrix (indeed - at equilibrium - it relates current to thermodynamic forces).

\end{enumerate}

Note that Eq.~\eqref{ness} means also $D=(A\sigma)^{symm}$ where, for
a generic matrix $M$, we define its symmetrized
$M^{symm}=(M+M^T)/2$. Therefore if all variables have the same parity,
the equilibrium condition state above reads $D=A\sigma$. This happens, for
instance, for overdamped Langevin equations, such as the one
considered here, with physical
interpretation~\eqref{PhysicalInterpretation}.

\subsection{The Response analysis} \label{sec:respmemory}

Thanks to linearity of equations (\ref{eqmotion}) the response properties
of the system can be easily calculated
\begin{equation}
{\bf R}(t)= e^{-At} \label{respformula}
\end{equation}
Where we have defined ${\bf R}(t)\equiv\overline{\frac{\partial
    X_{i}(t)}{\partial X_{j}(0)}}$.

Moreover, by a direct comparison between Eqs. (\ref{CorrOnsager}) and (\ref{respformula}) gives:
\begin{equation}
{\bf R}(t)= {\bf {C}}(t)\sigma^{-1}  \label{responsecorr}
\end{equation}
Where $\sigma^{-1}$ is the inverse matrix of (\ref{sigmamatrix}). 

By deriving with respect to time equation
(\ref{respformula}), and substituting into (\ref{responsecorr}) , one has
\begin{equation}
{\bf R}(t)={\bf \dot{C}}(t)(A\sigma)^{-1} \label{responseover}
\end{equation}
note that this is a particular case of a generalized
response equation, also called Generalized Fluctuation Dissipation
Relation (GFDR)~\cite{A72,FIV90,LCZ05,SS06,VBPV09,BMW09}.
For instance within the physical interpretation given in~\eqref{phys_ident}, the Onsager matrix reads
\begin{equation}
A\sigma= \left(
\begin{array}{cc}
\frac{T_{1}}{\gamma_{1}} & \Sigma \Delta T \\ 
 -\Sigma \Delta T & \frac{T_{2}}{\gamma_{2}}
\end{array}
\right)
\end{equation}
where $\Sigma=\frac{k}{(k+k_{2}) \gamma_{1}+(k+k_{1})\gamma_{2}}$ and,
as usual $\Delta T=(T_{1}-T_{2})$. Note that $A\sigma$ is diagonal if
$k=0$ or $\Delta T=0$.

For a correct comparison with  standard literature,
one must change slightly the definition of response used until this
moment. Let us suppose to make a perturbation of the Hamiltonian
(\ref{Hamiltonian}) with a term $-h(t)x_{1}$.  From equations of motion
(\ref{complete}) one has
\begin{equation}
\frac{\overline{\delta x_1}(t)}{\delta  h(0)}=\frac{1}{\gamma_{1}}\frac{\overline{\delta x_1}(t)}{\delta x_1(0)}.
\end{equation}
With such a mapping, one may write the linear response formula~\eqref{responseover} for the degree of freedom $x_1$ as
\begin{equation} \label{split}
\frac{\overline{\delta x_1}(t)}{\delta h(0)}=
\frac{(A\sigma)^{-1}_{11}}{\gamma_{1}}\frac{d}{dt}\langle
x_1(t)x_1(0)\rangle+\frac{(A\sigma)^{-1}_{12}}{\gamma_{1}}\frac{d}{dt}\langle
x_1(t)x_2(0)\rangle,
\end{equation}
where the two contributions on the right side to the response depends on the
time-scale of observation. 

When $T_{1}=T_{2}$ or $k=0$, $(A\sigma)_{12}$ vanishes and one recovers the
equilibrium condition $D=A\sigma$ (equivalent to reciprocal relations
for overdamped variables), together with the known equilibrium
fluctuation dissipation relation, $R_{ij}(t)\propto \dot {C_{ij}}(t)$.

\subsection{Entropy production}\label{sec:MarkEntr}

In this system it is easy to compute the entropy production functional of a
single trajectory.  Let us consider a general trajectory $\{ {\bf
  X}(s)\}_0^t$ and its time-reversal $\{ \mathcal{I}{\bf    X}(s)\}_0^t$. Lebowitz-Spohn defined the fluctuating entropy production
functional $W_t$ as follows~\cite{LS99}:
\begin{equation}
W_t'=\log\frac{\rho[{\bf X}(0)]P(\{ {\bf X}(s)\}_0^t)}{\rho[{\bf X}(t)]P(\{
   \mathcal{I}{\bf X}(s)\}_0^t)}=W_t+b_t \label{entropy_definition}
\end{equation}
with 
\begin{equation}
b_t=\log\{\rho[{\bf X}(0)]\}-\log\{\rho[({\bf X})]\}, \label{borderW}
\end{equation}
where $f[{\bf X}(0)]$ is the stationary distribution, i.e. the bivariate
Gaussian with covariance given by Eq. (\ref{sigmamatrix}) and
$P(\{ {\bf X}(s)\}_0^t)$ is the probability of the trajectory
introduced in equation (\ref{OnsagerMachlup}). Lebowitz and Sphon have
shown that the average (over the steady ensemble) of $W_{t}$, if
detailed balance is not satisfied, increases with time, while the term
$b_t$, usually known as ``border term'', is usually negligible for
large times, unless particular conditions of ``singularity''
occur~\cite{ZC03,PRV06,BGGZ06}.

For simplicity of notation, let us define
$\left(\begin{array}{c}F_{x}(\bX)\\F_{y}(\bX)\end{array}\right)\equiv
-A\bX$. In order to write down an explicit expression, it is necessary
to establish the behavior of the variables under time reversal
(e.g. positions are even and velocities are odd under time inversion
transformation). Let us assume that under time reversal it holds
$\overline{X_{i}}=\epsilon_{i}X_{i}$, where $\epsilon_{i}$ can be $+1$
or $-1$, using also $\ebX \equiv (\epsilon_1 X_1, \epsilon_2
X_2)$. Then one can define
\begin{eqnarray} 
F_i^{rev}(\bX)&=&\frac{1}{2}[F_i(\bX)-\epsilon_i F_i(\ebX)]=-\epsilon_iF_i^{rev}(\ebX) \label{drev}\\
F_i^{ir}(\bX)&=&\frac{1}{2}[F_i(\bX)+\epsilon_i F_i(\ebX)]=\epsilon_iF_i^{ir}(\ebX) \label{dirr}.
\end{eqnarray}
Given this notation~\cite{R89} it is possible to write down a compact
form for the entropy production\footnote{note that if the correlation
  matrix of the noise is not diagonal, this formula is slightly
  different.} simply by substituting equation (\ref{OnsagerMachlup})
into (\ref{entropy_definition}), obtaining:
\begin{equation} 
W_t = \sum_{k}D^{-1}_{kk}\int_{0}^{t}ds F^{ir}_{k}\left[ \dot{X}_{k}-F^{rev}_{k}\right] \label{formulone2}.
\end{equation}
Formula (\ref{formulone2}) is valid also in presence on non-linear
terms and with several variables. \\ From now on, in order to carry on
the calculations, it is necessary to take a decision on the parity of
the variables, under the time-reversal transformation. We will discuss
explicitly the overdamped dynamics case
(\ref{PhysicalInterpretation}), in which the variables $X_1$ and
$X_2$, being positions, are both even under the change of
time. Overdamped cases are usually simpler because the $F^{rev}$ terms
vanish. The non-overdamped case is discussed in~\ref{appinertia} and
has the same technical level with the difference that the velocity
variable is odd under time-reversal.  The exact expression of the
entropy production includes also border terms, which are not extensive
in time.  We do not include those terms in the calculations, since we
are interested in the asymptotic expression. \\ Using (\ref{formulone2}), the entropy
production is calculated to be
\begin{equation}
W_{t}=\frac{1}{D_1}\int_{0}^{t}dt' \lambda X_2\dot{X_1}-\alpha
X_1\dot{X_1}+ \frac{1}{D_2}\int_{0}^{t}dt' \mu X_1\dot{X_2} -\gamma
X_2\dot{X_2}\label{Entropy0}
\end{equation}
Note that the terms $\int dt X_1\dot{X_1}$ and $\int dt X_2\dot{X_2}$
are not extensive in time. Therefore the entropy production, for large
times, Eq. (\ref{Entropy0}) can be recast into
\begin{equation}
W_t\simeq
\left[\frac{\lambda}{D_1}-\frac{\mu}{D_2}\right]\int_{0}^{t}X_2\dot{X_1}dt'. \label{Entropyxx}
\end{equation}

It is possible to calculate the mean value of the entropy production rate (a limit for large times is meant)
\begin{eqnarray}
\frac{1}{t}\left<W_{t}\right>&\simeq&
\left[\frac{\lambda}{D_1}-\frac{\mu}{D_2}\right]\frac{1}{t}\int_{0}^{t}X_2\dot{X_1}dt'=\nonumber\\ 
&=&\left[\frac{\lambda}{D_1}-\frac{\mu}{D_2}\right]\left<X_2\dot{X_1}\right>. \label{EntropyMed}
\end{eqnarray}
Equation (\ref{EntropyMed}) can be closed by substituting the
equation of motion (\ref{eqmotion}) and the values of the static
correlations (\ref{sigmamatrix}), obtaining
\begin{equation}
\frac{1}{t}\left<W_{t}\right>=\frac{(D_2 \lambda -D_1 \mu )^2}{D_1 D_2 (\alpha +\gamma )} \label{entropy_production_mean}
\end{equation}

The formula applied to the physical interpretation~\eqref{phys_ident}, gives:
\begin{equation}
\frac{1}{t}\left<W_{t}\right>=
\frac{(k)^2}{(k+k_{1})\gamma_2+(k+k_{2})\gamma_1}\frac{\Delta T^2}{T_2T_1}. \label{entropy_production_mean_xx}
\end{equation}

It is immediate to recognize in formula
(\ref{entropy_production_mean_xx}) that the mean rate is always
positive, as expected. Moreover it is zero at equilibrium and in other
more trivial cases, namely when the dynamical coupling term $k$
goes to zero. It can approach to zero also in the limit of time scale
separation, but we will return on this point in Section
\ref{sec:effective}.


\section{Out-of-equilibrium information and cross-correlations} \label{sec:entropy_information}

In order to predict the response of an equilibrium system it is
sufficient to know its autocorrelation, as stated from the
fluctuation dissipation theorem.  In a broad sense, autocorrelation
and response have the same information content.  On the contrary we
have shown that the cross correlations between different degrees of
freedom plays a crucial role in non-equilibrium response.  The same
is true by considering the average entropy production rate: it is zero
at equilibrium because the cross correlation between $X_2$ and
$\dot{X}_1$ vanish.

In experiments or numerical simulations, however, if only $X_1$ is
observed, one is tempted to describe it by some effective stochastic
process which relegates the role of other degrees of freedom to some
kind of noise. The crudest way of doing it is neglecting any
time-delayed coupling of $X_1$ with other variables: of course such a
model is - in the absence of other external forces - necessarily an
equilibrium model, and cannot agree with observations; nevertheless,
the comparison with the equilibrium expectation can - in some cases -
lead to interesting interpretations. In the following we review the
case of effective temperatures, which are deduced by forcing a
comparison between non-equilibrium and equilibrium response
(autocorrelation), and in the case of extreme time-scale separation,
carry useful information about the two non-equilibrium
thermostats. After that, we also discuss a more {\em informed} way of
modelling the system, by considering time-delayed effects of the other
degrees of freedom in terms of memory and colored noise. The
predictions of such a model are much closer to observations, but we
show that crucial pieces of the puzzle are still missing.

\subsection{Comparison with a single variable, equilibrium, model}\label{sec:effective}

Extending what is certainly true at equilibrium, one may insist in
comparing response and correlation, by defining~\cite{CKP97,CR03}
\begin{equation}
T^{(AB)}_{eff}(t,t_{w})\equiv
\frac{R_{AB}(t,t_{w})}{\dot{C}_{AB}(t,t_{w})}, \label{FDT_RATIO_DEF}
\end{equation}
 with $\dot{C}_{AB}(t,t_{w})=\frac{d}{dt_{w}}\left< A(t) B(t_{w})\right>$, where $A$ and $B$ are two different observables of the system. The use
of two times $t_w$ and $t\ge t_w$ allows one to includes also cases
where the time translational invariance is not satisfied and
observables does depend in a non-trivial way on the waiting time
$t_{w}$ (for instance in aging systems). Equation
(\ref{FDT_RATIO_DEF}) represents an attempt to generalize the
temperature in system out of equilibrium, where ergodicity is broken.
The validity of a thermodynamic interpretation of this quantity is
clear in some limits, namely well separated
time-scales~\cite{CKP97,MPRR98,CK00,LN07}.

At a first sight, equation (\ref{FDT_RATIO_DEF})  appear in sharp contrast with the
``cross-correlation'' description given in Section \ref{sec:respmemory},
mainly because only the perturbed variable is involved~\cite{VBPV09}.  
Nevertheless in some cases also a partial view of the correlation response
plot is meaningful, in particular in the case of time scales
separation. For instance in the physical interpretation
(\ref{PhysicalInterpretation}), the model reveals an interesting and
non-trivial interplay of time-scales.  For simplicity let us consider
the case $k_{2}=0$.  A typical time for variable $x$, corresponding to
its relaxation time when decoupled by $y$ (i.e. $k=0$), is
$\tau_{1}=\frac{\gamma_{1}}{k_{1}}$. Analogously it is possible to
define a characteristic time for $y$: $\tau_{2}=\frac{\gamma_{2}}{k}$.
An interesting limit is the following:
\begin{eqnarray} \label{lssl}
\tau_{1}&\ll&\tau_{2}\nonumber\\
\frac{k}{k_1}&\sim& \frac{T_1}{T_2}, \nonumber
\end{eqnarray}
where the additional second condition guarantees that the interactions
have the same order of magnitude, so that the limit is non-trivial and
remains of pure non-equilibrium. In this case it can be shown that the
two timescales $\tau_1$ and $\tau_2$ correspond to those obtained by
inverting the two eigenvalues of the matrix $A$. Most importantly,
only in this limit the $FDT$ analysis of integrated response versus
correlation produces a two slopes curve, where $T_{1}$ and $T_{2}$ are
recognized as inverse of the measured slopes.
However this is a limit case, and more general conditions can
be considered.

In particular we consider the time-integrated response
$\chi_{11}=\int_0^t ds \frac{\overline{\delta x_1}(s)}{\delta h(0)}$,
and its two contributions appearing in the splitting formula~\eqref{split}
such that $\chi_{11}(t)=Q_{11}(t)+Q_{12}(t)$ and
\begin{eqnarray}
Q_{11}(t)&=\frac{(A\sigma)^{-1}_{11}}{\gamma_1}[C_{11}(0)-C_{11}(t)]\\
Q_{12}(t)&=\frac{(A\sigma)^{-1}_{21}}{\gamma_1}[C_{12}(0)-C_{12}(t)].
\end{eqnarray}

\begin{table} \begin{tabular}{|c|||c|c||c|c||c|c||c|c||c|c|}\hline
case &$T_2$ &$T_1$ &$\gamma_2$ &$\gamma_1$ &$\tau_2$ &$\tau_1$ &$k_{1}$
&$k$ &$\frac{1}{\lambda_-}$ &$\frac{1}{\lambda_+}$\\\hline 
a &2 &0.6 &200 &1 &200 &1 &1 &1 &400 &0.5 \\ 
b &5 &0.2 &20 &40 &30 &20 &2 &2/3 &47.3 &12.7\\ 
\hline 
\end{tabular} 
\caption{Table of parameters for the $2$ cases presented in
  Figures~\ref{fig:viol}. The effective time of
  the ``fast'' bath is defined as $\tau_1=\gamma_1/k_{1}$, while the
  relaxation time of particle $2$ is defined as $\tau_{2}=\gamma_2/k$
  ($k_{2}$ is always zero). The eigenvalues $\lambda_{1}$ and
  $\lambda_{2}$ of the dynamical matrix $A$ are also
  shown\label{paramtab}} 
\end{table}

Our choices of parameters, always with $T_1 \neq T_2$, are resumed in
Table~\ref{paramtab}: a case (a) where the time-scales are mixed, and
a case (b)  where scales are well separated.  Of course we do not
intend to exhaust all the possibilities of this rich model (given in more detail in~\cite{VBPV09}), but to
offer a few examples which may shed light on the role of cross correlations for linear response.

\begin{figure}
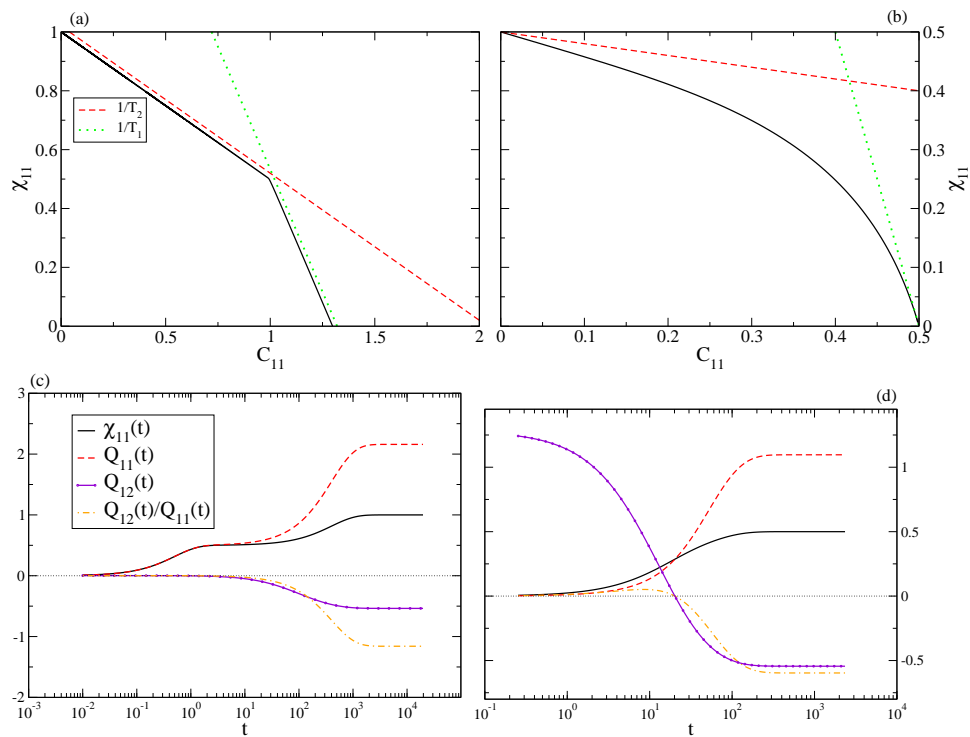

\includegraphics[width=6.3cm,clip=true]{violations_z_A.eps}
\includegraphics[width=6.3cm,clip=true]{violations_z_B.eps}\\
\includegraphics[width=6cm,clip=true]{tempi_A.eps}
\includegraphics[width=6cm,clip=true]{tempi_B.eps}
\caption{Frames a and b: parametric plots of integrated response
  $\chi_{11}(t)$ versus self-correlation $C_{11}(t)$ for the model in
  Eq.~\eqref{PhysicalInterpretation} with parameters given in
  Table~\ref{paramtab}. Lines with slopes equal to $1/T_2$, $1/T_1$
  are also shown for reference. Frames c and d: integrated response
  $\chi_{11}(t)$ as a function of time, for the same parameters. The
  curves $Q_{11}(t)$ and $Q_{12}(t)$, represent the two contributions
  to the response, i.e.  $\chi_{11}(t)=Q_{11}(t)+Q_{12}(t)$. The
  violet curve with small circles represents the ratio
  $Q_{12}(t)/Q_{11}(t)$.
\label{fig:viol}}
\end{figure}

The parametric plots, for the cases of Table~\ref{paramtab}, are shown
in Figure~\ref{fig:viol}, top frames. In the same figure, bottom frames, we present the
corresponding contributions $Q_{11}(t)$ and $Q_{12}(t)$ as
functions of time. We briefly discuss the two cases:
\begin{enumerate}

\item[(a)] In the ``glassy'' limit $\tau_2 \gg \tau_1$, with the
  constraint $y_0=\frac{T_1}{T_2}\frac{k_{1}}{k}\sim 1/2$, the well
  known broken line is found, see Fig.~\ref{fig:viol}a. Figure~\ref{fig:viol}c shows
  that $Q_{12}(t)$ is negligible during the first transient, up to the
  first plateau of $\chi(t)$, while it becomes relevant during the
  second rise of $\chi(t)$ toward the final plateau.

\item[(v)] If the timescales are not separated, the general form of
  the parametric plot, see Fig.~\ref{fig:viol}b, is a curve. In fact,
  as shown in Fig.~\ref{fig:viol}d, the cross term $Q_{12}(t)$ is
  relevant at all the time-scales. The slopes at the extremes of the
  parametric plot, which can be hard to measure in an experiment, are
  $1/T_1$ (at early times, high values of $C_{11}$) and some slope
  close to $1/T_2$ (at large times, low values of $C_{11}$). Apart
  from that, the main information of the parametric plot is to point
  out the relevance of the coupling of $x_{1}$ with the ``hidden''
  variable $x_{2}$.

\end{enumerate}

Note also that, if the relative coupling $k/k_1$ is changed, the
information on $T_1$ and $T_2$ may disappear from the
plot~\cite{VBPV09}. In summary the correct formula for the response is
always the GFDR: $\frac{\overline{\delta x}(t)}{\delta
  h(0)}=\dot{Q}_{11}+\dot{Q}_{12}$. However, the definition of an
effective temperature through the relation
$T_{eff}(t)\frac{\overline{\delta x}(t)}{\delta
  h(0)}=\dot{Q}_{11}(t)$, can be useful in those limits which are
relevant for glassy systems~\cite{CR03}, where the behavior of the additional term
$Q_{12}$ is such that $R \propto \dot{Q}_{11}$ in certain ranges of
time-scales.

\subsection[From the non-Markovian to the Markovian model]{Comparing with a single-variable non-equilibrium model with memory}
\label{sec:paradox}

Another classical approach to reduce the description of a many-body
system, e.g. to focus on a (possibly slow) single degree of freedom,
without losing the information of the reciprocal feedback between the
original variables, is to use a non-Markovian
description, with memory and colored noise.  In order to fix ideas,
let us consider again the linear model (\ref{eqmotion}). By
integrating formally the second equation one has
\begin{equation}
X_2(t)=\int_{-\infty}^t ds e^{-\gamma |t-s|}[\mu
  X_1(s)+\sqrt{2D_2}\phi_2(s)]. \label{int_aux}
\end{equation}
Putting (\ref{int_aux}) an equation for $X_{1}$ is obtained:
\begin{equation}
\dot{X}_1=-\alpha X_1 + \lambda \mu \int_{-\infty}^t ds e^{-\gamma
  |t-s|} X_1(s)+\eta(t)\label{NonMarkovianequation}
\end{equation}
with
\begin{equation}
\langle \eta(t)\eta(s) \rangle = 2D_1
\delta(t-s)+\frac{D_2\lambda^2}{\gamma}e^{-\gamma|t-s|}.
\end{equation}
it is worth noting that, with this mapping, the detailed balance
condition, given in the Markovian description by
$D_{1}\mu=D_{2}\lambda$, is ``translated'' into
\begin{equation}
\langle\eta(t)\eta(s)\rangle \propto e^{-\gamma|t-s|},
\end{equation}
which is the Fluctuation Dissipation Relation of the second kind, derived by Kubo for
generalized Langevin equations~\cite{KTH91}.

This mapping appear to be a harmless mathematical trick, and one is
tempted to consider the original system and the reduced model as equivalent. Actually it hides
 a loss of relevant information, detected for instance by
entropy production, as we discuss in the following.

The previous Section shows that if one takes the point of view of one
variable a different interpretation, respect to the two variables case,
of the ``violations''of the fluctuation response theorem can be
given. This interpretations are not in contrast each other, namely the
condition $T_{1}=T_{2}$ is always the ``equilibrium fingerprint''
which satisfies FDT. On the contrary, the scenario is different if one compare the
entropy production in the Non-Markovian system to what found in
Section \ref{sec:MarkEntr}.

Average entropy production for this non-Markovian model (originally
described in~\cite{ZBCK05}) is better studied  in frequency space, and
can be approached for a more general model. This is done with details in~\ref{appA}, while here we mention the main results. We start taking
into account the following one-dimensional Langevin equation
\begin{equation}
\label{eq:lang}
m\ddot{x}=-\gamma\dot{x} - h_x[x(t)] - \int_{-\infty}^{t} dt'\,
g(t-t')\, x(t') + \eta
\end{equation}
where $\eta(t)$ is Gaussian noise of zero mean and correlation
\begin{equation}
 \langle \eta(t)\, \eta(t')\rangle = \nu(t-t')
\end{equation}
with $\nu(t) = \nu(-t)$.
In this model one can calculate the path probability and its reversed.
The mean-value of Lebowitz-Sphon functional, ignoring all the
contributes non-extensive in time, reads
\begin{equation}
 \left\langle \log\frac{ {\cal P}\{x\} }{ {\cal P}\{\mathcal{I} x\} } \right\rangle
 = 
    -\int_{-\infty}^{+\infty} \frac{d\omega}{\pi}\,
    \omega\, 
    \left<\mbox{\rm Im}\bigl[x(\omega)h_x(-\omega)\bigr] \right>
     [\gamma + \psi(\omega)]\,
     \nu(\omega)^{-1}.\label{eq:epaMain}
\end{equation}
Where the average is performed on the space of trajectories.  The
functions appearing here are the Fourier transforms (See
\ref{app:NMentropyprod} for the details of the calculation). 

From equation (\ref{eq:epaMain}) it is easy to see that for the linear
case, namely for $h_{x}(x)\propto x(t)$ one has:
\begin{equation}
\left<\mbox{\rm Im}\bigl[x(\omega)h_x(-\omega)\bigr] \right>=0
\end{equation}
Remarkably, it predicts a vanishing entropy production also in the
case of the Linear model for $T_{1}\neq T_{2}$, in sharp contrast with
what found in (\ref{Entropyxx}) or in (\ref{EntropyVU}) for the case
of underdamped dynamics.

From this result it emerges that the two approaches represent the same
physical situation but with different level of details: moreover the
choice of the level of the description does not affect almost any of
the observables, for instance correlations and responses of the main
variable are unaffected, bringing the same FDT analysis of the models.
In order to focus on the reason of this difference, let us consider the
model with exponential memory (\ref{NonMarkovianequation}) that we rewrite here in a lightened
notation, for clarity:
\begin{equation}
\dot{x}= -h(x)+\lambda \mu \int_{t_{0}}^{t}e^{-\gamma(t-s)}x(s) + \eta(t) \equiv f_{x}+\eta(t) \label{NMmodel}
\end{equation}
The path probability  of this process, starting form the position $x_{0}$ at time $t_{0}$, can be expressed in the following form (see  \ref{appB} for details)

\begin{equation}
P[x| x_{0}]=\int \mathcal{D}\sigma \delta[\dot{x}+f_{x}+s(t)-\lambda\sqrt{2D_{y}}\int_{t_{0}}^{t}ds g(t-s)\phi_{y}(s)-
  \sqrt{2D_{x}}\phi_{x}(t)] \label{formula_crisanti}
\end{equation}
where we have used a simplified notation $\{ {\bf X}(s)\}_0^t\equiv
x$ and we have introduced $s(t)=-\lambda y_{0} g(t-t_{0})$. Moreover, 
$\mathcal{D}\sigma=dy_{0}
P_{0}(y_{0})\mathcal{D}\mu[\phi_{x}]\mathcal{D}\mu[\phi_{y}]$ where
the $\mu$'s are the Gaussian measures of the noises and $P_{0}$ is a
Gaussian distribution with zero mean and variance
$\frac{D_{y}}{\gamma}$.

After introducing an auxiliary process $\{ {\bf Y}(s)\}_0^t\equiv y$,
equation (\ref{formula_crisanti}) can be recast into:

\begin{eqnarray}
P[x| x_{0}]&=&\int \mathcal{D}\sigma
\mathcal{D}y
\delta[\dot{x}+h_{x}-\lambda y -
  \sqrt{2D_{x}}\phi_{x}(t)]\times\\
&&\delta[y - y_{0}g(t-t_{0})-\int_{t_{0}}^{t}ds g(t-s)[\mu x(s)+\sqrt{2D_{y}}\phi_{y}(s)]
\end{eqnarray}

After integrating over the noises, one obtains the following
expression for the probability
\begin{equation}
P[x| x_{0}]=\int dy_{0} P_{0}(y_{0})\int_{y(0)=y_{0}}
\mathcal{D}ye^{S(x,y)}
\end{equation}
where
\begin{equation}
S(x,y)=-\frac{1}{2D_{x}}\int_{t_{0}}^{t_{1}}dt[\dot{x}+h_{x}-\lambda y]^{2}-\frac{1}{2D_{y}}\int_{t_{0}}^{t_{1}}dt[\dot{y}+\gamma y-\mu x]^{2} \label{action2variable}
\end{equation}
It is straightforward to recognize that equation
(\ref{action2variable}) is the action of the corresponding two
variable stochastic process:

\begin{equation}
\left\{
\begin{array}{ccc}
\dot{x}&=& -h_{x}+\lambda y+\sqrt{2D_{x}}\phi_{x}\\
\dot{y}&=&-\gamma y+\mu x+\sqrt{2D_{y}}\phi_{y}
\end{array}
\right.
\end{equation}
for the particular choice of the initial condition $y_{0}$, following the Gaussian distribution $P_{0}$. 
This result shows how the path probability distribution of the model
(\ref{NMmodel}) is essentially given by a marginalization of the
corresponding Markovian one. From such an identification it is
straightforward to explain the results showed in the previous sections.

\subsection{General consequences of projections on entropy production}
If we denote with $\langle\dots \rangle_{x}$ the average over the
paths in the model (\ref{NMmodel}) and with $\langle\dots \rangle_{x,
  y}$ the average on the equivalent model on the auxiliary variable
one has \footnote{we omit to write down the border terms contributions for
simplicity.}, for an observable which depends only on $x$
\begin{equation}
\langle\mathcal{O}\rangle_{x}=\int\mathcal{D}x P(x)\mathcal{O}(x)=
\int\mathcal{D}x\mathcal{D}y e^{S(x,y)}\mathcal{O}(x)=\langle\mathcal{O}\rangle_{x, y}. \label{obsin2models}
\end{equation}
The relation (\ref{obsin2models}) show how, each observable of the
variable $x$ has the same values when computed in the two models. 

On the contrary 
\begin{equation}
\int\mathcal{D}x\mathcal{D}yP(x,y)\log\left[\frac{\int\mathcal{D}yP(x,y)}{\int\mathcal{D}y P_{1}(x,y)}\right]\neq 
\int\mathcal{D}x\mathcal{D}yP(x,y)\log\left[\frac{P(x,y)}{P_{1}(x,y)}\right].
\end{equation}
where we have defined with $P_{1}(x,y)$ the probability of the
inverted trajectory.

And, as a consequence, $\langle W \rangle_{x} \neq \langle W
\rangle_{x,y}$. This fact explains the difference observed. Moreover
it is simple to observe that
\begin{equation}
 \langle W \rangle_{x,y} -\langle W \rangle_{x}=\int \mathcal{D}x\mathcal{D}yP(x,y)\log{\frac{P(x,y)}{P_{1}(y\vert x)P(x)}}\geq 0 \label{RelativeEntropy}
\end{equation}
where the last inequality is a straightforward application of the
properties of Kullback-Leibler relative entropy, which is always
non-negative~\cite{KB51}. Then, this projection mechanism, in
general, has the effect of reducing entropy production. The equality
is satisfied if
\begin{equation}
P_{1}(y\vert x)=P(y\vert x) \label{current_loop}
\end{equation}
The physical meaning of~\eqref{current_loop} is clear: it represents a
sort of ``reduced'' detailed balance condition, it must be valid for
the variables one wants to remove from the description.\\ If one
removes from the description variables which are in equilibrium with 
respect to the others which remains, the procedure will not affect the
entropy production.  It is simple to note that this condition is not
valid, in general for the model (\ref{PhysicalInterpretation}), once
one decides to project away the variable $y$.
Under this point of view it is possible also to have an idea of why
the projection mechanism is not dangerous when the time scales are
well separated. Let us consider, for instance, the system in figure
\ref{fig1}. In the limit of $\tau_{2}\ll \tau_{1}$, the particle $1$
can be seen as blocked. Therefore the particle $2$ is in equilibrium
respect to the system ``thermostat + blocked particle $x_{1}$'' and eq
(\ref{current_loop}) is valid for every values of $x$.

\section{Conclusions and perspectives}
\label{sec:conclusion}

The linear equations~\eqref{eqmotion} constitute a simplified model of
a more complex, and perhaps realistic, system with $N \gg 1$ degrees of
freedom: such system $\Sigma$ is made of two sub-systems, say
$\Sigma_1$ and $\Sigma_2$, made of, respectively, $N_1$ and $N_2$
degrees of freedom, with $N_1+N_2=N$. The $N_i$ degrees of freedom of
sub-system $\Sigma_i$ are coupled to a thermostat at temperature $T_i$
and are immersed in an external confining potential, assumed harmonic
for simplicity. Furthermore, the $N_i$ degrees of freedom of sub-system
$\Sigma_i$ interact among themselves by intermolecular potential which
are, in general, not harmonic. In each subsystem $\Sigma_i$ there is
also a probe with position $x_i$ and momentum $p_i$, with mass
much larger than all the others in the same sub-system: such
condition on the masses of the probes is sufficient to
expect a linear Langevin-like dynamics for this degree of freedom,
where the (non-linear) interaction with all other molecules is
represented by an uncorrelated noise, while a linear velocity drag is due to
collisional relaxation, and of course the external harmonic potential
is still present, reproducing the situation of
Figure~\ref{fig1} and Eq.~\eqref{complete}. Finally, these two ``slow'' degrees of freedom
(with respect to the faster and lighter molecules) are
coupled one to the other by some potential $V(x_1-x_2)$. This coupling is the only
connection between systems $\Sigma_1$ and $\Sigma_2$. 

In the absence of the coupling between the probes, the two systems
remain separated and each one thermalizes to its own thermostat. When
the coupling is present, the whole system $\Sigma$ will have the
possibility to relax toward an overall equilibrium, but this is
prevented by the presence of the two thermostats which are ideally
infinite and never change their own temperature. The results is a
non-equilibrium steady state where energy is continuously transferred
on average from the hot to the cold reservoir. Such situation is quite
simple, but the nature of the coupling may pose some ambiguities when
the system is represented by the simplified $2$-variables
model. Indeed the above picture holds even if the coupling potential
$V(x_1-x_2) $ is harmonic: however in the harmonic case the modes at different
frequencies, i.e. $\tilde{X}_1(\omega),\tilde{X}_2(\omega)$ will be
decoupled. So, what is driving the system toward equilibrium,
i.e. exchanging heat or producing entropy? In the harmonic case, the
only channel for heat to flow is the one connecting
$\tilde{X}_1(\omega)$ to $\tilde{X}_2(\omega)$ with the same $\omega$:
the two components of the same mode are at different temperature and
can exchange heat. In summary, each mode has its own channel, which is
separated from the others. When the $2$-variables model is reduced to
the $1$-variable model with memory, the information about this channel
is completely lost because the two thermostats are reduced to only
one. Each ``cycle'' at frequency $\omega$ which behaves as a loop with
a given current, is flattened to a harmonic oscillator with zero net
current. The only remaining entropy production belongs to the exchange
between different modes. In this sense the single variable model does
not faithfully reproduce the full entropy production of the whole
system. On the other side, if some non-linearities are present, there are other
``channels'' of thermalization, due to the coupling between different
modes, even of the same variable: such channels are still active after
the projection to the single variable mode, and they continue to
contribute (maybe not exactly with the same average value) to a
non-zero entropy production.  In~\ref{example} we discuss an example
where two ``channels'' for entropy production are present (unbalance
of temperatures and an external force) and their different fates,
after a reduction of the description, is discussed.

\begin{figure}[h]
\begin{center}
\includegraphics[width=8cm,clip=true]{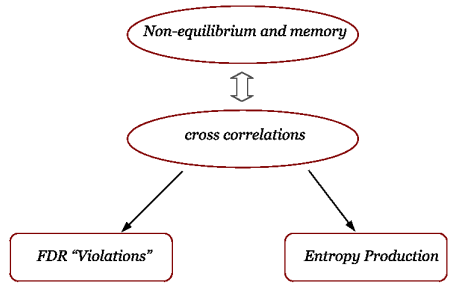}
\caption{ Non equilibrium and memory effects produce non-trivial
 correlations among different degrees of freedom. Taking into account
 this aspect, it is possible to explain pure non equilibrium
 phenomena like FDT violations and entropy production.  }
\end{center}
\end{figure}

This energy passing mechanism is evidently given by the correlations
between different degrees of freedom. Such a role is
crucial for two aspects:

\begin{itemize}
\item The response of the system to an impulsive perturbation is
  $R_{11} = a \langle \dot{X}_1(t)X_1(0)\rangle + b\langle \dot{X}_1(t)X_2(0)\rangle$ ,
  where a and b are some constants. As expected, for the equilibrium
  limit $(T_{1} \rightarrow T_{2})$, $b\rightarrow 0$ and the usual fluctuation
  response relation holds. On the contrary, when more than a
  thermostat is present, a coupling between different degrees of
  freedom emerges, “breaking” the usual form of the response relation.

\item The entropy production rate can be calculated by using the
  \emph{Onsager-Machlup} formalism. Also in this case, the rate is
  proportional to the cross correlations $\dot{X}_1X_2$ with a pre-factor depending
  on the two temperatures $T_{1}$ and $T_{2}$, and vanishing in the
  limit $T_{1}\rightarrow T_2$.

\end{itemize}

 These conclusions are not specific for the ``two variables'' model
 (\ref{eqmotion}). As mentioned before also other variable and some
 non-linearity can be inserted and the same description is still
 valid.

\appendix
\section{Generalized Langevin equations and non-equilibrium issues} \label{appA}

In this Appendix we study linear response and entropy production for a
particular generalized Langevin equation. Part of the results obtained
here have been obtained in similar or different ways in~\cite{ZBCK05}.

\subsection{Set up}
Consider the following simple one-dimensional Langevin equation
\begin{equation}
\label{eq:lang1}
\dot{x} = -h[x] + \eta
\end{equation}
where $\eta(t)$ is Gaussian noise of zero mean and correlation
\begin{equation}
 \langle \eta(t)\, \eta(t')\rangle = \nu(t-t')
\end{equation}
with $\nu(t) = \nu(-t)$.
The force term $h[x]$ contains a local in time part, denoted $h_x$, and a 
linear memory term, 
\begin{equation}
  h[x(t)] = h_x[x(t)] - \int_{-\infty}^{t} dt'\, g(t-t')\, x(t')
\end{equation}
Both $\nu(t)$ and $g(t)$ are left unspecified. 

Also we are interested into the stationary regime, so we let the initial time
to $-\infty$, and the final one to $+\infty$. Under this assumption the
probability of the a trajectory generated by the Langevin equation 
(\ref{eq:lang1}) is
\begin{equation}
\label{eq:pdt}
   {\cal P}\{x\} \propto \exp\left\{ 
    -\frac{1}{2}\int_{-\infty}^{+\infty} dt\,dt'\,
    \bigl[\dot{x}(t) + h[x(t)]\bigr] \nu^{-1}(t-t') 
    \bigl[\dot{x}(t') + h[x(t')]\bigr]
    \right\}
\end{equation}
where $\nu^{-1}(t)$ is the inverse of $\nu(t)$ defined as
\begin{equation}
  \int_{-\infty}^{+\infty} ds\, \nu(t-s)\,\nu^{-1}(s-t') = 
  \int_{-\infty}^{+\infty} ds\, \nu^{-1}(t-s)\,\nu(s-t') = \delta(t-t').
\end{equation}

By going in the Fourier space,
\begin{equation}
  x(t) = \int_{-\infty}^{+\infty}\frac{d\omega}{2\pi}\,
              e^{-i \omega t}\,x(\omega)
\quad  \longleftrightarrow \quad
  x(\omega) = \int_{-\infty}^{+\infty} dt \, e^{i\omega t}\,x(t).
\end{equation}
the probability (\ref{eq:pdt}) becomes
\begin{equation}
\label{eq:pdw}
   {\cal P}\{x\} \propto \exp\left\{ 
    -\frac{1}{2}\int_{-\infty}^{+\infty} \frac{d\omega}{2\pi}\,
    \bigl[-i\omega x(\omega) + h(\omega)\bigr] \nu(\omega)^{-1} 
    \bigl[ i\omega x(-\omega) + h(-\omega)\bigr]
    \right\}
\end{equation}
where $\nu^{-1}(\omega) = 1 / \nu(\omega)$, with $\nu(-\omega) = \nu(\omega)$,
and
\begin{equation}
  h(\omega) = \int_{-\infty}^{+\infty} dt \, e^{i\omega t}\, h[x(t)].
\end{equation}

\subsection{Entropy production}
\label{app:NMentropyprod}
Consider now the reversed trajectory $\mathcal{I} x(t) = x(-t)$.
Its probability follows from (\ref{eq:pdw}) by noticing that 
$\mathcal{I} x(\omega) = x(-\omega)$.
To compute the ratio between the probability of a trajectory $x$ and its
reversed $\mathcal{I} x$ we then have to separate the terms even and odd under 
the replacement  $x(\omega) \to x(-\omega)$ into (\ref{eq:pdw}).
To this end we have to look closer to $h(\omega)$.

From its definition we have
\begin{equation}
 h(\omega) = h_x(\omega) - 
   \int_{-\infty}^{+\infty} dt \, e^{i\omega t}\, 
   \int_{-\infty}^{t} dt'\, g(t-t')\, x(t')
\end{equation}
Now
\begin{eqnarray}
 \int_{-\infty}^{t} dt'\, g(t-t')\, x(t') &=& 
 \int_{-\infty}^{+\infty}\frac{d\omega}{2\pi}\, x(\omega)\,
 \int_{-\infty}^{t} dt'\, e^{-i \omega t'}\, g(t-t') 
\nonumber\\
  &=&
 \int_{-\infty}^{+\infty}\frac{d\omega}{2\pi}\, 
        e^{-i \omega t}\, x(\omega)\,
 \int_{0}^{\infty} dt'\, e^{i \omega t'}\, g(t')
\end{eqnarray}
so that 
\begin{equation}
 h(\omega) = h_x(\omega) - g(\omega)\, x(\omega)
\end{equation}
with
\begin{eqnarray}
 g(\omega) &=&  \int_{0}^{\infty} dt\, e^{i \omega t}\, g(t)
       \nonumber\\
           &=&   \int_{0}^{\infty} dt'\, \cos(\omega t)\, g(t)
             + i \int_{0}^{\infty} dt\, \sin(\omega t)\, g(t)
\nonumber\\
         &=& \phi(\omega) + i \omega\, \psi(\omega)
\end{eqnarray}
where
\begin{eqnarray}
  \phi(\omega) &=& \int_{0}^{\infty} dt'\, \cos(\omega t)\, g(t)
\\
  \psi(\omega) &=& \int_{0}^{\infty} dt\, \frac{\sin(\omega t)}\omega\, g(t)
\end{eqnarray}
are real even functions of $\omega$.
Collecting all terms we have
\begin{equation}
\label{eq:hw}
 h(\omega) = h_x(\omega) - \phi(\omega)\, x(\omega) -
 i\omega\,\psi(\omega)\,x(\omega)
\end{equation}
and (\ref{eq:pdw}) takes the form

\begin{eqnarray}
\label{eq:pdws}
   {\cal P}\{x\} &\propto& \exp\left\{ 
    -\frac{1}{2}\int_{-\infty}^{+\infty} \frac{d\omega}{2\pi}\,
    \bigl[-i\omega \widetilde{x}(\omega) + \widetilde{h}_x(\omega)\bigr] 
     \nu(\omega)^{-1} 
    \bigl[ i\omega \widetilde{x}(-\omega) + \widetilde{h}_x(-\omega)\bigr]
    \right\}
\nonumber \\
 &\propto&
\exp\Bigl\{ 
    -\frac{1}{2}\int_{-\infty}^{+\infty} \frac{d\omega}{2\pi}\,
    \bigl[\omega^2 \widetilde{x}(\omega)\widetilde{x}(-\omega) 
          + \widetilde{h}_x(\omega)\widetilde{h}_x(-\omega)\bigr] 
     \nu(\omega)^{-1} 
\nonumber\\
 &\phantom{\propto}&\hskip40pt
    +\frac{1}{2}\int_{-\infty}^{+\infty} \frac{d\omega}{2\pi}\,
    i\omega\,
    \bigl[\widetilde{x}(\omega)\widetilde{h}_x(-\omega) 
          - \widetilde{x}(-\omega)\widetilde{h}_x(\omega)\bigr] 
     \nu(\omega)^{-1} 
    \Bigr\}
\end{eqnarray}
where we have used the short-hand notation
\begin{equation}
 \widetilde{x}(\omega)   = x(\omega)   + \psi(\omega)\,x(\omega), \qquad
 \widetilde{h}_x(\omega) = h_x(\omega) - \phi(\omega)\,x(\omega).
\end{equation}
The first integral in the exponential is now even under the replacement 
$x(\omega) \to x(-\omega)$, while the second is odd. As a consequence the so 
called {\sl entropy production} reads:
\begin{eqnarray}
\label{eq:ep}
\log\frac{ {\cal P}\{x\} }{ {\cal P}\{\mathcal{I}x\} } &=&
    \int_{-\infty}^{+\infty} \frac{d\omega}{2\pi}\,
    i\omega\,
    \bigl[\widetilde{x}(\omega)\widetilde{h}_x(-\omega) 
          - \widetilde{x}(-\omega)\widetilde{h}_x(\omega)\bigr] 
     \nu(\omega)^{-1} 
\nonumber\\
 &=&
    -\int_{-\infty}^{+\infty} \frac{d\omega}{\pi}\,
    \omega\, 
    \mbox{\rm Im}\bigl[\widetilde{x}(\omega)\widetilde{h}_x(-\omega) 
                 \bigr] 
     \nu(\omega)^{-1} 
\nonumber\\
 &=&
    -\int_{-\infty}^{+\infty} \frac{d\omega}{\pi}\,
    \omega\, 
    \mbox{\rm Im}\bigl[x(\omega)h_x(-\omega)\bigr] 
     [1 + \psi(\omega)]\,
     \nu(\omega)^{-1} 
\end{eqnarray}
since the term proportional to $x(\omega)\,x(-\omega)$ is projected out
when one takes the imaginary part.
Note that if $h_x$ is linear in $x$ then the entropy production vanishes
for all trajectories.
Taking the average over all trajectories, weighted with (\ref{eq:pdw}), one
gets the average entropy production
\begin{equation}
\label{eq:epa}
 \left\langle \log\frac{ {\cal P}\{x\} }{ {\cal P}\{\mathcal{I}x\} } \right\rangle
 = 
    -\int_{-\infty}^{+\infty} \frac{d\omega}{\pi}\,
    \omega\, 
    \left<\mbox{\rm Im}\bigl[x(\omega)h_x(-\omega)\bigr] \right>
     [1 + \psi(\omega)]\,
     \nu(\omega)^{-1}.
\end{equation}

In a general non-equilibrium set up, the entropy production grows
linearly in time, namely
\begin{equation}
 \left\langle \log\frac{ {\cal P}\{x\} }{ {\cal P}\{\mathcal{I}x\} } \right\rangle
\sim  \sigma T \qquad \textrm{for } T\gg 1
\end{equation}

where $\sigma$ is the entropy production rate. Formally one can
introduce the mean entropy production rate:

\begin{equation}
 \left\langle \log\frac{ {\cal P}\{x\} }{ {\cal P}\{\mathcal{I}x\} }
 \right\rangle = \int_{-\infty}^{t} \sigma(s) d s
\end{equation}


By introducing the definition 
\begin{equation}
K(t)=\int_{-\infty}^{\infty}\frac{d\omega}{2\pi} [1 + \psi(\omega)]\,
     \nu(\omega)^{-1}
\end{equation}
and by exploiting the properties of the Fourier Transform and ignoring
sub leading terms, from (\ref{eq:epa}) one arrives to the following
identification:
\begin{equation}
\sigma_{t}=\int_{-\infty}^{t}dt' K(t-t')\bigl[\dot {x}(t)h_x(t')+\dot{x}(t')h_x(t)\bigr]\label{NMentropyrealspace}
\end{equation}
which coincides with the one derived in~\cite{ZBCK05}.

\subsection{Response analysis and fluctuation dissipation theorem}
The response $G(\omega)$ of the system to a (small) perturbation is
\begin{equation}
\label{eq:eg}
  G(\omega) = - \left. \frac{\delta}{\delta\,\epsilon(\omega)}
         \bigl\langle x(\omega) \bigr\rangle_{\epsilon}\right|_{\epsilon = 0}
\end{equation}
where the average is over all trajectories weighted with the probability 
(\ref{eq:pdw}) and $h(\omega) \to h(\omega) + \epsilon(\omega)$.
A simple calculation shows that
\begin{equation}
  G(\omega) = - \frac{\delta}{\delta\,h(\omega)}
         \bigl\langle x(\omega) \bigr\rangle
           = \nu(\omega)^{-1}\, 
              \bigl[ i\omega\, \langle x(\omega)\, x(-\omega)\rangle
                    + \langle x(\omega)\, h(-\omega)\rangle
             \bigr]
\end{equation}
where the averages are now computed with the unperturbed weight (\ref{eq:pdw}).
Note both $G(\omega)$ and the correlation function $C(\omega)$ 
are defined without the 
factor $2\pi$. This can be eliminated by a proper normalization of 
(\ref{eq:eg}). We do not do it to keep the equations lighter.

Using now (\ref{eq:hw}), and 
\begin{equation}
  C(\omega) = \langle x(\omega)\,x(-\omega)\rangle
\end{equation}
the response function $G(\omega)$ takes the form
\begin{equation}
\label{eq:Gw}
  G(\omega) = \nu(\omega)^{-1} \bigl[
                i\omega [1 + \psi(\omega)]\,C(\omega) 
                + \langle x(\omega)\, h_x(-\omega)\rangle
                - \phi(\omega)\,C(\omega)
             \bigr].
\end{equation}
Then 
\begin{equation}
 \mbox{\rm Im}\,G(\omega) = \nu(\omega)^{-1} \bigl[
                   [1 + \psi(\omega)]\, \omega\,C(\omega)
            + \mbox{\rm Im}\, \langle x(\omega)\, h_x(-\omega)\rangle
                \bigr]
\end{equation}
so that
\begin{equation}
\label{eq:betaw}
  \beta(\omega) = \frac{2\,\mbox{\rm Im}\,G(\omega)}{\omega\,C(\omega)} = 
     \frac{2}{\nu(\omega)} \left[ 
               1 + \psi(\omega) + 
           \frac{1}{\omega\,C(\omega)} 
            \mbox{\rm Im}\, \langle x(\omega)\, h_x(-\omega)\rangle
           \right]
\end{equation}
If FDT is obeyed $\beta(\omega)$ is constant and equal to the inverse of the 
temperature. For this reason $\beta(\omega)^{-1}$ has been called the FDT 
``effective'' temperature.
The FDT effective temperature is usually defined in the time domain as the 
ratio
\begin{equation}
  \beta(t) = - \frac{G(t)}{\partial_t C(t)}.
\end{equation}
However, if $\beta(\omega)$ depends on $\omega$, or $\beta(t)$ on $t$, then
$\beta(\omega)$ it is not the Fourier Transform of $\beta(t)$.

Since $\nu(\omega)$ is proportional to the temperature $T$, a better definition
would be
\begin{eqnarray}
\label{eq:mw}
  m(\omega) = T \beta(\omega) &=& 
   \frac{2\,T\,\mbox{\rm Im}\,G(\omega)}{\omega\,C(\omega)} = \nonumber\\ 
     &=& \frac{2T}{\nu(\omega)} \left[ 
               1 + \psi(\omega) + 
           \frac{1}{\omega\,C(\omega)} 
            \mbox{\rm Im}\, \langle x(\omega)\, h_x(-\omega)\rangle
           \right],
\end{eqnarray}
and similarly
\begin{equation}
  m(t) = T\beta(t) = - \frac{T\,G(t)}{\partial_t C(t)}.
\end{equation}
The non-dimensional functions $m(\omega)$ and $m(t)$ give indications on
the violation of the FDT, both in the frequency and time domains.

The vanishing of the (average) entropy production is commonly taken as a 
signature of equilibrium. By comparing Eq. (\ref{eq:epa}) with 
(\ref{eq:betaw}), or (\ref{eq:mw}), we see that this does not necessarily 
imply that the FDT is obeyed. Indeed in this case, one gets 
\begin{equation}
\label{eq:ratio}
\frac{2\,\mbox{\rm Im}\,G(\omega)}{\omega\,C(\omega)} = 
     \frac{2}{\nu(\omega)} \left[ 
               1 + \psi(\omega)
           \right]
\end{equation}
that, in general, is function of $\omega$. This is, for example, the case
of linear $h_x$, as it will be illustrated in the Section \ref{sec:linear}.

From (\ref{eq:ratio}) we see that there are two contributions to the ratio.
The first, $\nu(\omega)$, from the noise, and the second, $\psi(\omega)$, 
from the deterministic part of the equation. Tracking back the calculation we
see that the latter follows from the term $-i\omega\,\psi(\omega)\,x(\omega)$ 
in $h(\omega)$, see Eq. (\ref{eq:hw}). Transforming this contribution back to
time space, we have
\begin{eqnarray}
\label{eq:iwpsi}
 -i\omega\,\psi(\omega)\,x(\omega) &\Rightarrow& 
  \frac{d}{dt}\,\int_{-\infty}^{+\infty}\frac{d\omega}{2\pi}\,
       e^{-i\omega t}\,\psi(\omega)\,x(\omega)
\nonumber\\
&\Rightarrow& 
      \int_{-\infty}^{+\infty}dt'\, \frac{d}{dt}\psi(t-t')\,x(t')
\end{eqnarray}
where
\begin{equation}
  \psi(t) = \int_{-\infty}^{+\infty}\frac{d\omega}{2\pi}\,
              e^{-i \omega t}\, \psi(\omega)
= \int_{-\infty}^{+\infty}\frac{d\omega}{2\pi}\,
              e^{-i \omega t}\, 
       \int_{0}^{\infty} dt'\, \frac{\sin(\omega t')}\omega\, g(t')
\end{equation}


\subsection{An example: the linear case} \label{sec:linear}
To illustrate the general results of previous sections we shall consider
the special case of 
\begin{equation}
\label{eq:example}
 h_x[x] = -\alpha x, \quad
  g(t) = \lambda \mu e^{-\gamma |t|}, \quad
  \nu(t) = 2 D_x \delta(t) + \frac{\lambda^2 D_y}{\gamma}\,e^{-\gamma |t|}.
\end{equation}
for which we have:
\begin{equation}
 \nu(\omega) = 2D_x \frac{\omega^2 + r^2}{\omega^2 + \gamma^2}
\end{equation}
where
 $r^2 = \gamma^2 + \lambda^2 D_y / D_x$,
\begin{equation}
  h(\omega) = \left[\alpha - \frac{\lambda\mu}{\gamma - i\omega}\right]\,
        x(\omega) 
\end{equation}
and
\begin{equation}
\label{eq:phipsi}
  \phi(\omega) = \lambda\mu\frac{\gamma}{\omega^2 + \gamma^2},
\qquad
  \psi(\omega) = \lambda\mu\frac{1}{\omega^2 + \gamma^2}.
\end{equation}

These expression are obtained from the coupled linear Langevin equations
\begin{eqnarray}
 \dot{x} &=& -\alpha x + \lambda y + \sqrt{2 D_x}\,\phi_x \nonumber\\
 \dot{y} &=& -\gamma y + \mu     x + \sqrt{2 D_y}\,\phi_y 
\end{eqnarray}
where $\phi_x(t)$ and $\phi_y(t)$ are two independent white noises, 
by eliminating $y$.
Solving the second equation for $y(t)$, and inserting the result into the
first equation, one indeed ends up with 
\begin{equation}
 \dot{x} = -\alpha x 
           + \lambda \mu \int_{-\infty}^{t} dt'\, e^{-\gamma(t-t')}\,x(t')
           + \eta
\end{equation}
with the ``noise''
\begin{equation}
 \eta(t) = \sqrt{2D_x}\,\phi_x(t) 
      + \lambda\sqrt{2D_y} \int_{-\infty}^tdt'\, e^{-\gamma(t-t')}\,\phi_y(t')
\end{equation}
from which (\ref{eq:example}) follow.

The entropy production (\ref{eq:ep}), and hence the average entropy production,
vanishes for this system. This result is not unexpected since
the probability of a trajectory is Gaussian:
\begin{equation}
\label{eq:pdlw}
   {\cal P}\{x\} \propto \exp\left\{ 
    -\frac{1}{2}\int_{-\infty}^{+\infty} \frac{d\omega}{2\pi}\,
    x(\omega)\,C(\omega)^{-1}\, x(-\omega)
    \right\}
\end{equation}
with 
\begin{equation}
\label{eq:Cwl}
 C(\omega) = 
    \left| \frac
                    {\gamma -i\omega}
                    {(\alpha -i\omega)(\gamma - i\omega) -\mu\lambda}
             \right|^2\,\nu(\omega)
\end{equation}
The system is linear
\begin{equation}
\label{eq:xwl}
  x(\omega) = G(\omega)\, \eta(\omega)
\end{equation}
then 
$C(\omega) = G(\omega)\,\nu(\omega)\,G(-\omega)$.
Inserting this expression into (\ref{eq:Gw}):
\begin{equation}
 G(\omega) = \nu(\omega)^{-1}\left[ i\omega  + \alpha 
                       - \frac{\mu\lambda}{\gamma + i\omega}
                        \right]\, G(\omega)\,\nu(\omega)\,G(-\omega)
\end{equation}
a simple calculation leads to
\begin{equation}
\label{eq:Gwl}
  G(\omega) =  \frac
                    {\gamma -i\omega}
                    {(\alpha -i\omega)(\gamma - i\omega) -\mu\lambda}
\end{equation}
in agreement with (\ref{eq:Cwl}) (and (\ref{eq:xwl})).

From this expression, or directly from (\ref{eq:ratio}), one then obtains
\begin{equation}
 \beta(\omega) = \frac{2}{\nu(\omega)}\left[
            1 + \lambda\mu\frac{1}{\omega^2 + \gamma^2}
                        \right]
              = \frac{1}{D_x} \frac{\omega^2 + \gamma^2 + \lambda\mu}
                                   {\omega^2 + \gamma^2 + \lambda^2D_y/D_x}.
\end{equation}
From this expression we see that FDT is obeyed if and only if
\begin{equation}
  \lambda\mu = \lambda^2 D_y/D_x \quad \longrightarrow \quad
  \mu D_x = \lambda D_y
\end{equation}
In the general case we can identify the two limiting regimes
\begin{eqnarray}
\beta(\omega) &\simeq& \frac{1}{D_x} \qquad |\omega| \gg 1
\\
\beta(\omega) &\simeq& \frac{1}{D_x} \frac{\gamma^2 + \lambda\mu}
                                   {\gamma^2 + \lambda^2D_y/D_x}.
\qquad |\omega| \sim 0
\end{eqnarray}
Finally from the expression (\ref{eq:phipsi}) it follows
$\psi(t) = \lambda\mu \exp(-\gamma |t|)/2\gamma$, so that 
in this case
\begin{equation}
  -i\omega\,\psi(\omega)\,x(\omega) \Rightarrow
     -\frac{\lambda\mu}{2} \int_{-\infty}^{+\infty}dt'\, 
     \mbox{\rm sign}(t-t')\, e^{-\gamma |t-t'|}\, x(t').
\end{equation}
This term gives informations on the ``asymmetry'' forward/past of the 
trajectories around time $t$, measured on a  time interval $|t-t'|$ ruled 
by the memory characteristic decay time $\gamma^{-1}$.

It is instructive to compare the previous results with those obtained from the
study of the model in the time domain. To this end we observe that $G(\omega)$ 
and $C(\omega)$, Eqs. (\ref{eq:Gwl}) and (\ref{eq:Cwl}), can be written as
\begin{equation}
 G(\omega) = \frac{\gamma - i\omega}
         {(\omega + i\omega_{-})(\omega + i\omega_{+})}
\end{equation}
\begin{equation}
 C(\omega) = \nu(\omega) \frac{\gamma^2 + \omega^2}
         {(\omega^2 + \omega_{-}^2)(\omega^2 + \omega_{+}^2)}
\end{equation}
where
\begin{equation}
 \omega_{\pm} = \frac{1}{2}\left[\alpha + \gamma \pm \sqrt{-\Delta}\right]
\end{equation}
with
\begin{equation}
-\Delta = (\alpha + \gamma)^2 - 4(\alpha\gamma - \lambda\mu)
        = (\alpha - \gamma)^2 + 4\lambda\mu.
\end{equation}
Causality, i.e. the requirement $G(t) = 0$ if $t < 0$,  implies 
$\mbox{\rm Re}\,\omega_{\pm} > 0$, which in turns leads to the constraint
$\alpha\gamma - \lambda\mu > 0$. The sign of $-\Delta$ can be positive or
negative, so we may have both purely exponential functions or modulated 
exponential functions.

Taking the inverse Fourier Transform of $G(\omega)$ and $C(\omega)$ we have
\begin{equation}
\label{eq:Gtl}
 G(t) = \frac{\theta(t)}{\sqrt{-\Delta}}
     \left[ (\omega_{+} - \gamma)\,e^{-\omega_{+}t}
            -(\omega_{-} - \gamma)\,e^{-\omega_{-}t}
        \right]
\end{equation}
and
\begin{equation}
\label{eq:Ctl}
 C(t) = \frac{1}{2(\alpha+\gamma)\sqrt{-\Delta}}
      \left[ 
 \frac{\omega_{+}^2 - \gamma^2}{\omega_{+}}\,\nu(i\omega_{+})\,e^{-\omega_{+}t}
-
\frac{\omega_{-}^2 - \gamma^2}{\omega_{-}}\,\nu(i\omega_{-})\,e^{-\omega_{-}t}
        \right]
\end{equation}
so that
\begin{equation}
\beta(t) = 2(\alpha+\gamma) \frac
              {(\omega_{+} - \gamma)\,e^{-\omega_{+}t}
                -(\omega_{-} - \gamma)\,e^{-\omega_{-}t}
              }
              {
     (\omega_{+}^2 - \gamma^2)\,\nu(i\omega_{+})\,e^{-\omega_{+}t}
-
     (\omega_{-}^2 - \gamma^2)\,\nu(i\omega_{-})\,e^{-\omega_{-}t}
              }
\end{equation}
The limiting values of $\beta(t)$ for the case of real 
$\omega_{\pm}$ are
\begin{eqnarray}
  \beta(t) &\simeq& \frac{1}{D_x} \qquad t \sim 0^+
\\
 \beta(t) &\simeq& \frac{1}{D_x} \frac{\gamma^2 + \lambda\mu - \omega_{-}^2}
                                   {\gamma^2 + \lambda^2D_y/D_x - \omega_{-}^2}
\qquad t \gg 1
\end{eqnarray}

\section {A model with inertia}
\label{appinertia}

In some cases it is helpful to consider an alternative interpretation
of Eqs.~\eqref{eqmotion}: this is realized, for instance, when
considering a massive granular intruder in a gas of other granular
particles driven by a stochastic external energy injection. Indeed the
steady state dynamics of the intruder velocity $V \equiv X_1$ is
fairly modelled by the following equation~\cite{SVCP10}
\begin{eqnarray}
M\dot{V}=-\Gamma(V-U)+\sqrt{2\Gamma T_g}\phi_1\\ \nonumber
M'\dot{U}=-\Gamma' U - \Gamma V+\sqrt{2 \Gamma' T_b} \phi_2,  \label{PhysicalInterpretation2}
\end{eqnarray}
where $M$ is the intruder mass, $\Gamma$ is the drag coefficient of
the surrounding granular fluid, $T_g$ is the granular temperature of
the fluid, $U$ is related to a {\em local average force field} of particles
colliding with the intruder, $M'$ and $\Gamma'$ are two parameters
which characterize the effective mass and drag of the auxiliary field
$U$, and finally $T_b$ is the temperature of the external bath which
keeps steady the system. This model is cast to model (\ref{eqmotion})
by identifying $X_1 \to V$, $X_2 \to U$ and mapping the parameters in
the following way:
\begin{equation} \label{phys_ident2}
\alpha \to \frac{\Gamma}{M} \;\;\; \lambda \to \frac{\Gamma}{M}
\;\;\; \gamma \to \frac{\Gamma'}{M'} \;\;\; \mu \to
-\frac{\Gamma}{M'} \;\;\; D_1 \to \frac{\Gamma T_g}{M^2} \;\;\;
D_2 \to \frac{\Gamma' T_b}{(M')^2}.
\end{equation}
Note that in this interpretation the main variable $X_1$ is a velocity
and therefore is odd under time-reversal, while it was even in the
overdamped case described in Section \ref{sec:MarkEntr}; note also
that $\mu$ here is negative.  This model possess the same mathematical
structure of the overdamped case, and can be treated in a similar
manner. Also even if the two systems model different physical
contests, the underlying non-equilibrium mechanism is evidently the
same: the presence of a more than one bath, which put the system out
of equilibrium.

The stationary state for the case under scrutiny is characterized by
the following matrix of covariances:
\begin{equation}
\sigma=\left(
\begin{array}{cc}
\frac{T}{M}+\Theta \Delta T & \Theta \Delta T \\
\Theta \Delta T & \frac{T_{1}}{M'}+\frac{\Gamma}{\Gamma'}\Theta \Delta T \\
\end{array}
\right) \label{SigmaNESS}
\end{equation}
where we have introduced
$\Theta=\frac{\Gamma\Gamma'}{(\Gamma+\Gamma')(M'\Gamma+M\Gamma')}$ and
$\Delta T = T_{b}-T_{g}$. From this reparametrization emerges that,
when $T_{b}=T_{g}$ the two variables are uncorrelated. 

Linear response for the variable $V$ reads:
\begin{equation}
\frac{\delta V(t)}{\delta V(0)}=\sigma^{-1}_{VV}\langle
V(t)V(0)\rangle + \sigma^{-1}_{UV}\langle
V(t)U(0)\rangle \label{respunderd}
\end{equation}
In the equilibrium case when $T_{g}=T_{b}$, one has that the Einstein
relation is recovered. This is quite simple to observe, since
$\sigma^{-1}_{UV}=0$ and one has $R_{VV}(t)\equiv C_{VV}(t)$. 
In the general case $\sigma_{UV}\neq 0$ and the mobility $\mu$ is not
simply given by the integral of the autocorrelation of velocity. This
apparent ``violation'' is restored only if all the couplings between
the different degrees of freedom are taken into account.

\begin{figure}[h] 
\begin{center} 
\includegraphics[width=8cm,clip=true]{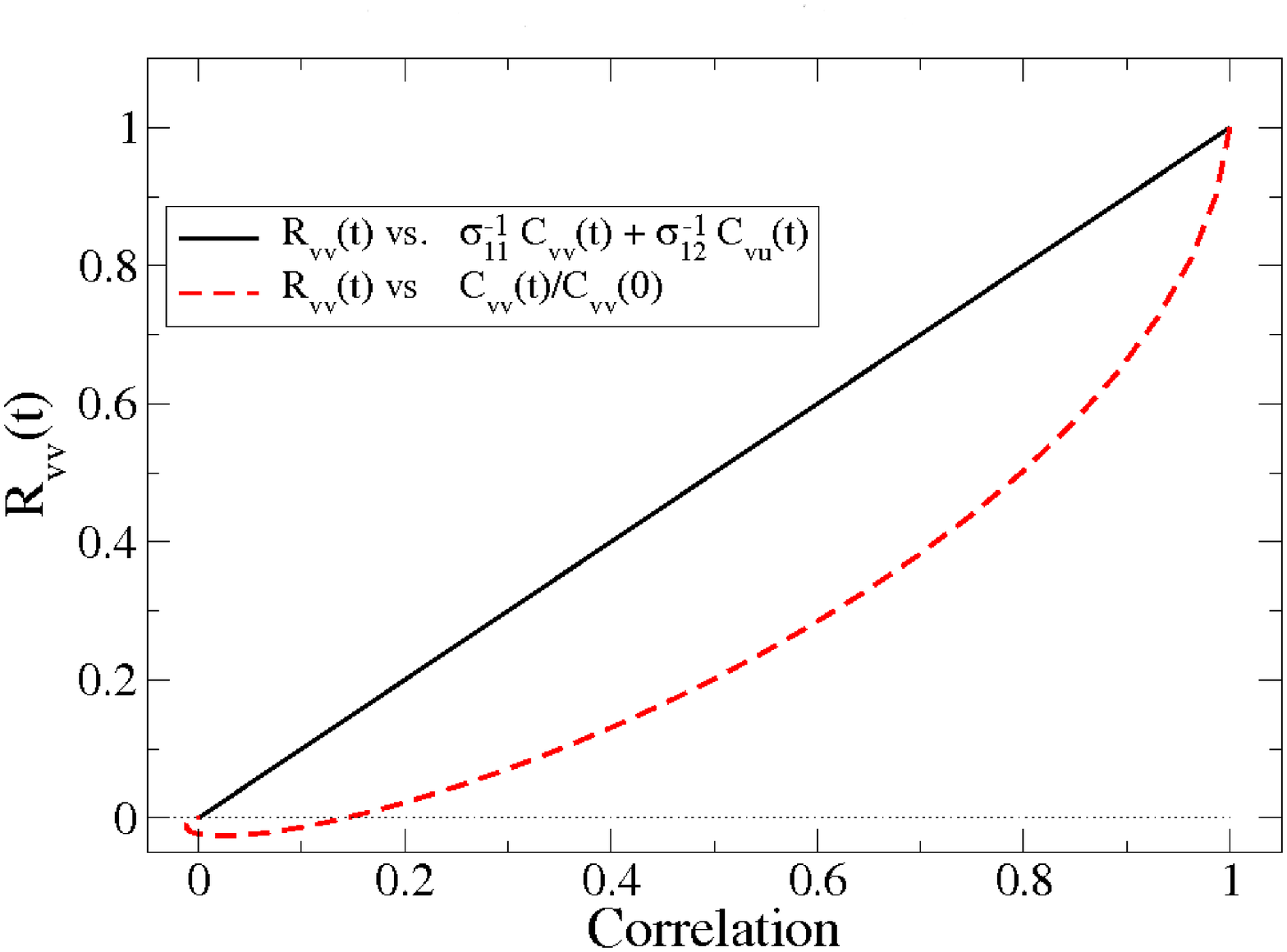}
\caption{ Response of variable $V$ in function of the unperturbed
  autocorrelation (red dashed line), when $T_{g}\neq T_{b}$.
Linearity is restored if one use the GFDR.  \label{fig:kubo} }
\end{center}
\end{figure}

This difference is emphasized in Fig.~\ref{fig:kubo}, where the
Einstein Relation is violated and the GFDR holds: response
$R_{VV}(t)$, when plotted against $C_{VV}(t)$, shows a non-linear
relation. Anyway a simple linear plot is restored when the response is
plotted against the linear combination of correlations indicated by
formula~\eqref{respunderd}. In this case it is evident that the
``violation'' can not be interpreted by means of any effective
temperature, namely from this plot is not possible to extract the two
temperature underlying the dynamics. In analogy of what discussed in
Section \ref{sec:respmemory}, the non linear relation between correlation and
response is a consequence of having ``missed'' the coupling between
variables $V$ and $U$, which gives an additive contribution to the
response of $V$.

 For instance, let us take the limit $M'\gg 1$ in
 (\ref{PhysicalInterpretation2}). In this limit, the relaxing time
 $\tau_{U}$ of the variable $U$ diverges. Roughly speaking, for time
 intervals lower than $\tau_{U}$, one can consider $U(t)$ equal to the
 initial value $U_{0}$. Within this approximation the equation of the
 particle becomes:
\begin{equation}
M\dot{V}=-\Gamma(V-U_{0})+\sqrt{2\Gamma T_g}\phi_1,
\end{equation}
which is clearly the equation of a particle moving in a time
independent force field. The response function is easily calculated:
\begin{equation}
R_{VV}=\frac{M}{T_{g}}\langle
V(t)[V(0)-U_{0}]\rangle. \label{response_constant_field}
\end{equation}
An interpretation of (\ref{response_constant_field}) is evident: it
represent a sort of Einstein Relation, between mobility and diffusion,
calculated in the ``Lagrangian frame'' of the particle. Thanks to the
time scale separation between the two variables, one can see that a linear
relation is valid,  apart from a slight modification. The
time-scales separation restores the linearity and a Response
Correlation plot is meaningful. Note that this restoring of the
fluctuation dissipation relation in a Lagrangian frame has attracted
some authors, and has inspired several works, both
theoretical~\cite{AS12,CG09} and experimental~\cite{GPCCG09,GPCM11}. 

For the entropy production, a procedure identical to the overdamped
case can be performed also in the case with inertia, taking care of
the parity transformations $V\rightarrow -V$, $U\rightarrow U$ under
time reversal operation. In this case one obtains for the general
model
\begin{equation}
W_t\simeq\left[ \frac{\alpha\lambda}{D_1}+\frac{\mu\gamma}{D_2}\right]\int_{0}^{t}X_1X_2dt'. \label{Entropy1}
\end{equation}
Making the substitution (\ref{phys_ident2}):
\begin{equation}
W_t\simeq \Gamma \left(\frac{1}{T_{g}}-\frac{1}{T_{b}}\right)\int_{0}^{t}V(t')U(t')dt'. \label{EntropyVU}
\end{equation}
and its mean entropy production rate is 
\begin{equation}
\frac{1}{t}\left<W_{t}\right>=\frac{\Gamma\Theta(\Delta
  T)^2}{T_{g}T_{b}} \label{entropy_production_mean_UV}
\end{equation}

\section{How to generate time translational invariant colored noise} \label{appB}

Let us suppose that our purpose is to reproduce equation with memory
(as, for instance in equation (\ref{NMmodel})). By exploiting the idea
described in section \ref{sec:entropy_information}, one can starts from the Markovian problem:
\begin{eqnarray}
 \dot{x} &=& -\alpha x + \lambda y + \sqrt{2 D_x}\,\xi_x \nonumber\\
 \dot{y} &=& -\gamma y + \mu     x + \sqrt{2 D_y}\,\xi_y  \label{app:markov}
\end{eqnarray}

However, once integrated the second equation of (\ref{app:markov}) , starting
from time $t_{0}$ with initial condition $y_{0}$ and substituted it in
the first one, the expression for the \emph{effective} noise of the variable $x$ is  

\begin{equation}
\eta(t)=\lambda y_{0}g(t-t_{0})+\lambda\sqrt{2D_{y}}\int_{t_{0}}^{t}ds
g(t-s)\phi_{y}(s)+ \sqrt{2D_{x}}\phi_{x}(t). \label{app:effnoise}
\end{equation}
At a first sight it seems that Eq. (\ref{app:effnoise}) does not
satisfy a time translational condition, namely
$\langle\eta(t)\eta(t')\rangle \neq f(t-t') $. In order to show this, let us
write down the probability distribution of the noise:
\begin{equation}
P[\eta |y_{0}]=\int \mathcal{D}\sigma \delta[\eta(t)-s(t)-\lambda\sqrt{2D_{y}}\int_{t_{0}}^{t}ds g(t-s)\phi_{y}(s)-
  \sqrt{2D_{x}}\phi_{x}(t)] 
\end{equation}
where $s(t)=\lambda y_{0} g(t-t_{0})$ takes into account the dependence of the initial condition of the variable $y$. By introducing the \emph{hat} variable $\hat{\eta}(t)$ and by exploiting the integral representation of the delta function, one has
\begin{equation}
P[\eta |y_{0}]=\int\mathcal{D}\hat{\eta}\mathcal{D}\phi_{x}\mathcal{D}\phi_{y}A_{\eta}B_{\phi_{x}}C_{\phi_{y}}
\end{equation}
Where we have introduced the following notations:
\begin{eqnarray}
A_{\eta}&=& \exp\left\{\int_{t_{0}}^{t_{1}}dt\phantom{.}i\hat{\eta}(t)
  [\eta(t)-\lambda
    y_{0}g(t-t_{0})]\right\}\\
B_{\phi_{x}}&=& \exp\left\{-i\sqrt{2D_{x}}\int_{t_{0}}^{t_{1}}dt
\hat{\eta}(t)\phi_{x}(t)\right\}\\
C_{\phi_{y}}&=&\exp\left\{-i\lambda\sqrt{2D_{y}}\int_{t_{0}}^{t_{1}}dt
\int_{t_{0}}^{t_{1}}dt' \hat{\eta}(t')g(t'-t)\phi_{y}(t)\right\}
 \end{eqnarray}

Now we use the identity $\langle e^{\lambda x} \rangle =
e^{\frac{1}{2}\lambda^{2}\langle x^{2}\rangle}$, which is valid for
Gaussian integrals, obtaining

\begin{eqnarray}
\int\mathcal{D}\phi_{x}
P[\phi_{x}]B_{\phi_{x}} &=& \exp\left\{-D_{x}\int_{t_{0}}^{t_{1}}\hat{\eta}^{2}(t)\right\}\\
\int\mathcal{D}\phi_{y}
P[\phi_{y}] C_{\phi_{y}}
 &=&  \exp\left\{-\lambda^{2}D_{y}\int_{t_{0}}^{t_{1}}dt
\int_{t_{0}}^{t_{1}}dt'\hat{\eta}(t)\Delta(t,t')\hat{\eta}(t')\right\}
\end{eqnarray}

where we have introduced
\begin{eqnarray}
  \Delta(t,t') & \equiv &\int_{t_{0}}^{t_{1}}dt' g(t-t'')g(t'-t'') \\
                &= & \frac{1}{2\gamma}\left[e^{-\gamma \vert t - t' \vert}-e^{-\gamma \vert t + t' - 2t_{0} \vert}\right]
\end{eqnarray}

Finally, by integrating over the $\hat{\eta}$ the following
\emph{Onsager-Machlup} probability distribution is obtained

\begin{equation}
P[\eta |y_{0}]=\exp\left\{-\frac{1}{2}\int_{t_{0}}^{t_{1}}dt
  \int_{t_{0}}^{t_{1}}dt'F[\eta(t)]\nu(t,t')F[\eta(t')]\right\}
\end{equation}
with 
\begin{eqnarray}
F[\eta(t)]&\equiv & \eta(t)-\lambda y_{0}g(t-t_{0}) \\
  \nu^{-1}(t,t')& =& 2D_{x}\delta(t-t')+\frac{\lambda^{2}D_{y}}{\gamma}
  \left[e^{-\gamma \vert t - t' \vert}-e^{-\gamma ( t + t' - 2 t_{0}
      )} \right]. \label{appeq:colored}
\end{eqnarray}

As expected, expression (\ref{appeq:colored}) is not of the requested
form: the autocorrelation is not time translational invariant and
dependence of the initial condition is explicit.  However, one can
choose the  initial condition $y_{0}$ randomly
with distribution $P_{0}$. Then, the final expression for the
distribution of the noise is obtained by integrating over the initial
condition:

\begin{equation}
P[\eta]=\int dy_{0}P_{0}(y_{0})P[\eta |y_{0}]
\end{equation}

By choosing $P_{0}$ of the Gaussian form with zero mean and variance
$\sigma^{2}=\frac{D_{y}}{\gamma}$, after some calculations, one obtains
a colored Gaussian process whose correlation is time translational
invariant, namely

\begin{equation}
P[\eta]=\exp\left\{-\frac{1}{2}\int_{t_{0}}^{t_{1}}dt
  \int_{t_{0}}^{t_{1}}dt'\eta(t)\nu(t,t')\eta(t')\right\}
\end{equation}
with
\begin{equation}
  \nu^{-1}(t,t')=2D_{x}\delta(t-t')+\frac{\lambda^{2}D_{y}}{\gamma}
 e^{-\gamma \vert t - t' \vert} \label{app:corr}
\end{equation}

In conclusion, from this example we learn the correct procedure to
reproduce a colored noise with correlation (\ref{app:corr}) by using an
auxiliary variable. In order to obtain it, it is necessary to choose
the initial condition $y_{0}$ from a specific random distribution.

\section{Many channels for entropy production: an example}
\label{example}

To make clear the discussion done in the conclusions, let us consider
a particular example where two different ``channels'' for entropy
production can be put in evidence. The example consists in a particle
subject to a non-equilibrium bath and to an external driving force
$F$: both mechanisms produce entropy. The velocity follows the
equation:
\begin{equation}
\dot{v}=-\int_{-\infty}^{t}\Gamma(t-t')v(t')dt'+F+\eta(t)
\end{equation}
where
\begin{equation}
\Gamma(t-t')=2\gamma_{f}\delta(t-t')+\frac{\gamma_{s}}{\tau_{s}} e^{-\frac{(t-t')}{\tau_{s}}}
\end{equation}
\begin{equation}
\langle\eta(t)\eta(t')\rangle=2T_f \gamma_{f}\delta(t-t')+ T_s\frac{\gamma_{s}}{\tau_{s}}e^{-\frac{\vert t-t'\vert}{\tau_{s}}}
\end{equation}
Clearly, due to the presence of the non conservative force $F$, the
particle reaches a non-zero average velocity:
\begin{equation}
V_{lim}=\frac{F}{\int_{0}^{+\infty}\Gamma(t)dt}=\frac{F}{\gamma_{s}+\gamma_{f}}.
\end{equation}
In this case, also in the non-Markovian description, an entropy
production rate does exist. Following formula
(\ref{NMentropyrealspace}) of the \ref{app:NMentropyprod}, such rate
reads\footnote{Note that formula (\ref{NMentropyrealspace}) has been
  derived in the overdamped case, however it is simple to verify that
  it is does not change in the case with inertia }:
\begin{equation}
\sigma^{diss}(t)=F\int_{-\infty}^{t}dt'K(t-t')\left[v(t)+v(t')\right]
\end{equation}
where
\begin{equation}
K(t)=\frac{\delta(t)}{T_{f}}+\frac{\gamma_{s}}{2\Omega
  T_{f}\gamma_{f}\tau^{2}_{s}}\left(1-\frac{T_{s}}{T_{f}}\right)e^{-\Omega\vert
  t \vert}\nonumber
\end{equation}
\begin{equation}
\Omega=\frac{1}{\tau_{s}}\sqrt{\frac{\gamma_{f} T_{f}+\gamma_{s} T_{s}}{\gamma_{f} T_{f}}}\nonumber
\end{equation}

The average entropy production rate can be exactly calculated,
yielding to the following result:
\begin{eqnarray}
\overline{\sigma^{diss}}&=& \left(\frac{1}{T_{f}}+\frac{
  \gamma_{s}}{T_{f}\gamma_{f}\tau^{2}_{s}\Omega^{2}}\left(1-\frac{T_{s}}{T_{f}}\right)\right)V_{lim}
F\nonumber \\ &=&\left(\frac{1}{T_{f}}+\frac{
  \gamma_{s}T_{s}}{\gamma_{s} T_{s}+\gamma_{f}
  T_{f}}\left(\frac{1}{T_{s}}-\frac{1}{T_{f}}\right)\right)V_{lim}F\label{NMentropyprod}
\end{eqnarray}
Such result clearly shows that the entropy production vanishes if the
external driving $F$ is removed; on the other side if $F \neq 0$ a
production exists even if $T_f=T_s$.

The same calculation for the mean entropy production rate can be
carried out also for the corresponding \emph{Markovian} system,
obtaining:
\begin{equation}
\overline{\sigma_{M}^{diss}}=\left(\frac{F}{T_{f}}\left<v\right>+ \gamma_{s}\left(\frac{1}{T_{s}}-\frac{1}{T_{f}}\right)\left<vu\right>\right)
\end{equation}
Note that, because of the driving force, the variable has non-zero
mean. Therefore, by using $\left<vu\right>= \left<\delta v\delta
u\right>+\left<u\right>\left<v\right>$, one obtains the following
result for the average entropy production rate:
\begin{equation}
\overline{\sigma_{M}^{diss}}=\frac{(T_{f}-T_{s})^{2}\gamma_{f}\gamma_{s}}{(\gamma_{s}+\gamma_{f})(1+\gamma_{f}\tau_{s})T_{f}T_{s}}
+\left(\frac{1}{T_{f}}+ \frac{\gamma_{s}}{\gamma_{s}+\gamma_{f}}\left(\frac{1}{T_{s}}-\frac{1}{T_{f}}\right)\right)V_{lim}F\label{em}
\end{equation}
The first term in the sum~\eqref{em} is completely absent in the
non-Markovian approach, Eq.~\eqref{NMentropyprod}, and is different
from zero even if $F=0$. The second term is slightly different from
(\ref{NMentropyprod}), where a weighted average on the temperatures is
present in the prefactor. They become identical when $\gamma_s \gg
\gamma_f$.
\clearpage
\section*{References}

\bibliographystyle{unsrt}
\bibliography{paper.bib}

\begin{thebibliography}{10}

\bibitem{M71}
J.C. Maxwell.
\newblock {\em Theory of Heat}.
\newblock Dover (New York), 1871, reprinted 2001.

\bibitem{S29}
L~Szilard.
\newblock \"uber die {E}ntropieverminderung in einem thermodynamischen {System
  bei Eingriffen intelligenter Wesen}.
\newblock {\em Z. Phys.}, 53:840, 1929.

\bibitem{L61}
R~Landauer.
\newblock Irreversibility and heat generation in the computing process.
\newblock {\em IBM J. Res. Dev.}, 5:183, 1961.

\bibitem{SU10}
T.~Sagawa and M.~Ueda.
\newblock Generalized jarzynski equality under nonequilibrium feedback control.
\newblock {\em Physical review letters}, 104(9):90602, 2010.

\bibitem{AS12}
David Abreu and Udo Seifert.
\newblock Thermodynamics of genuine nonequilibrium states under feedback
  control.
\newblock {\em Phys. Rev. Lett.}, 108:030601, Jan 2012.

\bibitem{C12}
S.~Ciliberto.
\newblock Private communication.
\newblock 2012.

\bibitem{KPB07}
R.~Kawai, JMR Parrondo, and C.V. den Broeck.
\newblock Dissipation: The phase-space perspective.
\newblock {\em Physical review letters}, 98(8):80602, 2007.

\bibitem{RJ07}
S.~Rahav and C.~Jarzynski.
\newblock Fluctuation relations and coarse-graining.
\newblock {\em Journal of Statistical Mechanics: Theory and Experiment},
  2007:P09012, 2007.

\bibitem{PPRV10}
A.~Puglisi, S.~Pigolotti, L.~Rondoni, and A.~Vulpiani.
\newblock Entropy production and coarse graining in markov processes.
\newblock {\em Journal of Statistical Mechanics: Theory and Experiment},
  2010:P05015, 2010.

\bibitem{CK00}
L~F Cugliandolo and J~Kurchan.
\newblock A scenario for the dynamics in the small entropy production limit.
\newblock {\em J Phys Soc Jpn}, 69:247, 2000.

\bibitem{VBPV09}
D~Villamaina, A~Baldassarri, A~Puglisi, and A~Vulpiani.
\newblock Fluctuation dissipation relation: how to compare correlation
  functions and responses?
\newblock {\em J. Stat. Mech.}, page P07024, 2009.

\bibitem{ZBCK05}
F~Zamponi, F~Bonetto, L~F Cugliandolo, and J~Kurchan.
\newblock A fluctuation theorem for non-equilibrium relaxational systems driven
  by external forces.
\newblock {\em J. Stat. Mech.}, page P09013, 2005.

\bibitem{SVGP10}
A~Sarracino, D~Villamaina, G~Gradenigo, and A~Puglisi.
\newblock Irreversible dynamics of a massive intruder in dense granular fluids.
\newblock {\em Europhys. Lett.}, 92:34001, 2010.

\bibitem{V06}
P~Visco.
\newblock Work fluctations for a {B}rownian particle between two thermostats.
\newblock {\em J. Stat. Mech.}, page P06006, 2006.

\bibitem{FI11}
H.C. Fogedby and A.~Imparato.
\newblock A bound particle coupled to two thermostats.
\newblock {\em Journal of Statistical Mechanics: Theory and Experiment},
  2011:P05015, 2011.

\bibitem{PV09}
A~Puglisi and D~Villamaina.
\newblock Irreversible effects of memory.
\newblock {\em Europhys. Lett.}, 88:30004, 2009.

\bibitem{KTH91}
R~Kubo, M~Toda, and N~Hashitsume.
\newblock {\em Statistical physics II: Nonequilibrium stastical mechanics}.
\newblock Springer, 1991.

\bibitem{CKP97}
L~F Cugliandolo, J~Kurchan, and L~Peliti.
\newblock Energy flow, partial equilibration, and effective temperatures in
  systems with slow dynamics.
\newblock {\em Phys. Rev. E}, 55:3898, 1997.

\bibitem{CR03}
A~Crisanti and F~Ritort.
\newblock Violation of the fluctuation-dissipation theorem in glassy systems:
  basic notions and the numerical evidence.
\newblock {\em J. Phys. A}, 36:R181, 2003.

\bibitem{R89}
H~Risken.
\newblock {\em The Fokker-Planck equation: Methods of solution and
  applications}.
\newblock Springer- {V}erlag, Berlin, 1989.

\bibitem{MSR73}
P~C Martin, E~D Siggia, and H~A Rose.
\newblock Statistical dynamics of classical systems.
\newblock {\em Phys. Rev. A}, 9:423, 1973.

\bibitem{OM53}
L~Onsager and S~Machlup.
\newblock Fluctuations and irreversible processes.
\newblock {\em Phys. Rev.}, 91:1505, 1953.

\bibitem{A72}
G~S Agarwal.
\newblock Fluctuation-disipation theorems for systems in non-thermal
  equilibrium and applications.
\newblock {\em Z. Physik}, 252:25, 1972.

\bibitem{FIV90}
M~Falcioni, S~Isola, and A~Vulpiani.
\newblock Correlation functions and relaxation properties in chaotic dynamics
  and statistical mechanics.
\newblock {\em Physics {L}etters {A}}, 144:341, 1990.

\bibitem{LCZ05}
E.~Lippiello, F.~Corberi, and M.~Zannetti.
\newblock Off-equilibrium generalization of the fluctuation dissipation theorem
  for ising spins and measurement of the linear response function.
\newblock {\em Physical Review E}, 71(3):036104, 2005.

\bibitem{SS06}
T~Speck and U~Seifert.
\newblock Restoring a fluctuation-dissipation theorem in a nonequilibrium
  steady state.
\newblock {\em Europhys. Lett.}, 74:391, 2006.

\bibitem{BMW09}
M.~Baiesi, C.~Maes, and B.~Wynants.
\newblock Fluctuations and response of nonequilibrium states.
\newblock {\em Physical review letters}, 103(1):10602, 2009.

\bibitem{LS99}
J~L Lebowitz and H~Spohn.
\newblock A {G}allavotti-{C}ohen-type symmetry in the large deviation
  functional for stochastic dynamics.
\newblock {\em J. Stat. Phys.}, 95:333, 1999.

\bibitem{ZC03}
R~van Zon and E~G~D Cohen.
\newblock Extension of the fluctuation theorem.
\newblock {\em Phys. Rev. Lett.}, 91:110601, 2003.

\bibitem{PRV06}
A~Puglisi, L~Rondoni, and A~Vulpiani.
\newblock Relevance of initial and final conditions for the fluctuation
  relation in markov processes.
\newblock {\em J. Stat. Mech.}, page P08010, 2006.

\bibitem{BGGZ06}
F~Bonetto, G~Gallavotti, A~Giuliani, and F~Zamponi.
\newblock Chaotic hypothesis, fluctuation theorem, singularities.
\newblock {\em J. Stat. Phys.}, 123:39, 2006.

\bibitem{MPRR98}
E~Marinari, G~Parisi, F~Ricci-Tersenghi, and JJ~Ruiz-Lorenzo.
\newblock Violation of the fluctuation-dissipation theorem in
  finite-dimensional spin glasses.
\newblock {\em J. Phys. A.}, 31:2611, 1998.

\bibitem{LN07}
L~Leuzzi and Th.~M Nieuwenhuizen.
\newblock {\em Thermodynamics of the Glassy State}.
\newblock Taylor {$\&$} Francis, 2007.

\bibitem{KB51}
S.~Kullback and R.A. Leibler.
\newblock On information and sufficiency.
\newblock {\em The Annals of Mathematical Statistics}, 22(1):79--86, 1951.

\bibitem{SVCP10}
A~Sarracino, D~Villamaina, G~Costantini, and A~Puglisi.
\newblock Granular brownian motion.
\newblock {\em J. Stat. Mech.}, page P04013, 2010.

\bibitem{CG09}
R.~Chetrite and K.~Gawedzki.
\newblock Eulerian and lagrangian pictures of non-equilibrium diffusions.
\newblock {\em Journal of Statistical Physics}, 137(5):890--916, 2009.

\bibitem{GPCCG09}
J.R. Gomez-Solano, A.~Petrosyan, S.~Ciliberto, R.~Chetrite, and K.~Gawedzki.
\newblock Experimental verification of a modified fluctuation-dissipation
  relation for a micron-sized particle in a nonequilibrium steady state.
\newblock {\em Physical review letters}, 103(4):40601, 2009.

\bibitem{GPCM11}
J.R. Gomez-Solano, A.~Petrosyan, S.~Ciliberto, and C.~Maes.
\newblock Fluctuations and response in a non-equilibrium micron-sized system.
\newblock {\em Journal of Statistical Mechanics: Theory and Experiment},
  2011:P01008, 2011.

\end{thebibliography}

\end{document}